\documentclass[runningheads,citeauthoryear]{apinv}
\usepackage{cite,graphics}
\usepackage[utf8]{inputenc}
\usepackage{graphicx}
\usepackage{marvosym}
\usepackage{url}
\usepackage{txfonts}
\usepackage{multirow}
\usepackage{longtable}
\usepackage{adjustbox}
\usepackage{tabularx}
\usepackage{caption}
\usepackage{xcolor}

\newcommand{\degree}{$\,^o\,$}

\newcommand{\rsun}{$\,R_\odot\,$}

\setkeys{Gin}{draft=False} %if True, all figures will print as just an empty box (useful when you don't want to wait for an entire dissertation with figures to compile)

\begin{document}

\title{Characterization of the Early Dynamics of Solar Coronal Bright Fronts}
\titlerunning{Characterization of Coronal...}

\author{Mohamed Nedal\thanks{Corresponding author} \and
        Kamen Kozarev \and
        Rositsa Miteva \and
        Oleg Stepanyuk \and
        Momchil Dechev
        }
\authorrunning{M. Nedal et al.}
% Command tocautor{} is used by the Latex to give author names 
% to the Contents of the volume (automatically generated)
\institute{Institute of Astronomy and National Astronomical Observatory - Bulgarian Academy of Sciences, 1784 Sofia, Bulgaria
\newline
\email{mnedal@nao-rozhen.org} % Just the corresponding author
}
\papertype{Submitted on xx.02.2024; Accepted on xx.xx.xxxx}	
% Papertype can be "Research report", "Review", "Invited lecture", "Conference talk", 
% "Conference poster", "Lecture at scientific seminar", "Summary of dissertation",  etc.

\maketitle

\begin{abstract}
We present a comprehensive characterization of 26 Coronal Mass Ejection (CME)-driven compressive waves known as Coronal Bright Fronts (CBFs) observed in the low solar corona between 2010 and 2017. These CBFs have been found to be associated with Solar Energetic Particle (SEP) events near Earth, indicating their importance in understanding space weather phenomena.
% aims heading
The aim of this study is to analyze and describe the early dynamics of CBFs using a physics-based heliospheric SEP forecasting system known as the Solar Particle Radiation Environment Analysis and Forecasting - Acceleration and Scattering Transport (SPREAdFAST) framework. This framework utilizes a chain of data-driven analytic and numerical models to predict SEP fluxes at multiple locations in the inner heliosphere by considering their acceleration at CMEs near the Sun and subsequent interplanetary transport.
% methods heading
To estimate the time-dependent plasma and compression parameters of the CBFs, we utilized sequences of base-difference images obtained from the Atmospheric Imaging Assembly (AIA) instrument on board the Solar Dynamics Observatory (SDO) satellite, and measurements of the height-time profiles of the CMEs obtained from the Large Angle and Spectrometric COronagraph (LASCO) instrument on board the Solar and Heliospheric Observatory (SOHO) satellite. We employed kinematic measurements and plasma model results to derive these parameters. The SPREAdFAST framework facilitated the analysis and correlation of these observations with SEP events near Earth.
% results heading
Our analysis yielded statistical relations and distributions for both the shocks and plasma parameters associated with the 26 CBFs investigated. By combining the observations from the AIA and LASCO instruments, as well as the data products from the SPREAdFAST framework, we obtained a comprehensive understanding of the early dynamics of CBFs, including their temporal evolution, plasma properties, and compressional characteristics. These findings contribute to the growing body of knowledge in the field and have implications for space weather forecasting and the study of SEP events.
\end{abstract}

\section*{Introduction}
Coronal bright fronts (CBFs), also known as extreme ultraviolet (EUV) waves, are disturbances that propagate over significant portions of the solar disk and off the solar limb. These waves can reach speeds faster than the local characteristic speed in the solar corona, transforming into shock waves. They are primarily driven by Coronal Mass Ejections (CMEs) or solar flares \cite{thompson_1999a, veronig_2010, vrsnak_2008, magdaleni_2010, nindos_2011}. In both radio and white-light observations, CBFs often appear as dome-shaped structures moving at speeds on the order of several hundred \kms \cite{pick_2006, nindos_2008, thompson_1998, thompson_1999b}. These structures consist of piled-up plasma with higher density, making them appear brighter in white-light images.

Observing and studying coronal shock waves remotely is typically done through EUV observations using space-based instruments such as the Atmospheric Imaging Assembly (AIA) on the Solar Dynamics Observatory (SDO) spacecraft \cite{lemen_2011}. Alternatively, shock waves can be indirectly observed through the detection of type II radio bursts, which are commonly associated with shock waves in the solar corona \cite{vrsnak_2008}.

The AIA instrument has provided valuable insights into the dynamics of the low solar corona over the past decade, thanks to its exceptional spatial and temporal resolution. Equipped with telescopes observing the solar disk in bands 193 and 211~\AA, the AIA instrument has demonstrated its ability to distinguish compressive waves in the lower corona \cite{patsourakos_2010, ma_2011, kozarev_2011}. These observations offer valuable information about the kinematics and geometric structure of CBFs. To accurately study the evolution of the wave's leading front, observations off the solar limb are preferred to mitigate projection effects, which may introduce ambiguities in estimating time-dependent positions and the global structure of the wave \cite{kozarev_2015}.

In situ observations of shock waves have revealed their classification into quasi-parallel, quasi-perpendicular, sub-critical, and super-critical shocks based on the angle between the wavefront normal vector and the upstream magnetic field lines \cite{tsurutani_1985}. Quasi-parallel shocks have a shock-field angle ($\theta_{BN}$) smaller than 45\degree, while quasi-perpendicular shocks have $\theta_{BN}$ greater than 45\degree. Supercritical shocks, often associated with accelerated particles, are promising candidates for generating type II radio bursts \cite{benz_1988}. However, obtaining accurate estimates of shock strength and obliquity solely from remote observations is challenging.

Recent studies have further elucidated the characteristics of CBFs both on the solar disk and off the limb, confirming their wave-like nature \cite{nitta_2013, long_2011, olmedo_2012}. Coronagraph observations, such as those obtained from the LASCO instrument on board the SOHO spacecraft \cite{domingo_1995}, have extended the investigation of shock waves beyond 2.5\rsun \cite{vourlidas_2003}, while EUV observations have provided evidence linking CMEs and EUV waves \cite{patsourakos_20l9}. Nevertheless, the appearance of shock waves in EUV observations is not yet fully understood \cite{kozarev_2011}. Emission measure modeling using the EUV channels of the AIA instrument allows for the estimation of temperature and density changes in the wavefront's sheath \cite{kozarev_2011}.

By employing multi-wavelength studies from the SOHO/LASCO and SDO/AIA instruments, valuable information about the relationship between white-light coronagraph observations and EUV observations of CMEs has been uncovered, shedding light on the properties of CBFs closer to the Sun \cite{warmuth_2015}. Factors such as the presence of nearby active regions or coronal holes can distort the initial morphological shape of CBFs \cite{ofman_2002, mann_2003, piantschitsch_2018}, and a connection between CBFs and chromospheric disturbances known as Moreton waves has been established \cite{thompson_1999b}.

Coronal shock waves are recognized as particle accelerators; however, the mechanisms through which Solar Energetic Particles (SEPs) are produced by coronal shocks throughout the inner heliosphere remain uncertain. Previous statistical studies have focused on comparing CME properties with those of SEP events \cite{kahler_1999, reames_1999, kahler_2001, kahler_2005, richardson_2015, papaioannou_2016}. Correlations between CME speeds and peak intensities of associated SEPs have been investigated \cite{kahler_2001}, and a relationship has been found between integrated SEP energy and CME energy \cite{kahler_2013}. Nonetheless, these correlations exhibit considerable scatter in the plots of CME speeds versus SEP fluxes \cite{kouloumvakos_2019}. Various factors may contribute to this scattering, including projection effects on shock parameters, inappropriate parameter choices, seed particle populations, the geometric structure of shocks, or the possibility that shocks may not be the primary particle accelerators \cite{kouloumvakos_2019}.

Many SEP events detected near Earth are not directly associated with Earth-detected shocks, indicating that these SEPs are likely accelerated much closer to the corona, possibly by shock waves \cite{reiner_2007}. In this study, we focus on 26 CBF events up to $\sim$17\rsun by combining observations and modeling tools from the Solar Particle Radiation Environment Analysis and Forecasting--Acceleration and Scattering Transport framework \cite[SPREAdFAST]{kozarev_2022}. Our aim is to estimate the ambient plasma interactions with CBFs and gain insights into the acceleration and transport of SEPs along the wavefronts of coronal shocks and compressive waves.

%We have completed full modeling of 26 events using the SPREAdFAST model chain, making this Sun-to-Earth study unparalleled in terms of the number of events modeled and the direct comparison between modeled in-situ fluxes and observations. Among these events, four have exhibited modeling difficulties and will be subjected to further investigation. Another third of the events, totaling 32, will be used for out-of-sample validation of forecasting. These events either did not show measurable shock waves using our method of kinematics measurements or had source active regions located well within the solar disk, making them impractical for our measurement method. However, they will still be modeled for comparison, utilizing time-dependent shape/kinematics of shock waves and diffusion coefficients obtained from tabulated values based on the other 26 events. The modeling chain will remain consistent for all events. All events in our study exhibit increases in SEP fluxes at 1 AU.
The paper is organized as follows: Section~\ref{s_data_methods} provides details about the data resources and analysis methods employed. The results are presented in Sections~\ref{s_case_study} and~\ref{s_stat_study}, accompanied by thorough discussions. Finally, Section~\ref{s_conclusions} concludes the study and provides a summary of the findings.

%%________________________________________________________________
\section{Data Analysis and Methods}
\label{s_data_methods}

\subsection{Observations}
We collected data from the SOHO/ERNE instrument, focusing on proton events within the energy range of 17-22 MeV, from 2010 to 2017. Initially, we obtained a total of 216 events. However, we applied several criteria to filter the data and arrive at the final list of events for this study.
Firstly, we excluded 39 proton events that were not associated with EUV waves and had no identified CMEs or flares. Additionally, 72 proton events were excluded because they lacked EUV wave associations, despite having identified CMEs/flares. We also removed 6 events with uncertain EUV waves from our analysis. Furthermore, 37 events were discarded due to immeasurable EUV waves.
Moreover, 36 events did not show measurable shock waves using our method of kinematics measurements. As a result, we proceeded with 26 events that exhibited measurable CBFs, allowing us to analyze them using our framework.
To initiate the analysis, we utilized image sequences obtained from the EUV channel 193 $\AA$ of the AIA instrument. These images had a 24-second cadence and served as the primary input for the SPREAdFAST framework.

The 26 selected events (Table~\ref{table_1}) were previously presented in our previous work \cite{kozarev_2022}. Table~\ref{table_1} provides details about the date of the CBF events, the start and end times of associated flares along with their class, and the source locations on the solar disk in helioprojective Cartesian coordinates. These coordinates were obtained from the Heliophysics Events Knowledge (HEK) database\footnote{HEK Database: \url{www.lmsal.com/isolsearch}}.

% all events had type III radio bursts.
\begin{table} % updated!
\caption{List of the CBF events with their associated flares and CMEs.}
\label{table_1}
%\small % Adjust font size as needed
%\footnotesize
\tiny
\setlength{\tabcolsep}{7pt} % Adjust column separation
\renewcommand{\arraystretch}{1.5} % Adjust row height
\begin{tabularx}{\textwidth}{*{12}{>{\centering\arraybackslash}X}}
    \hline 
    \textbf{ID} & \textbf{Event Date} & \textbf{Flare Start (UT)} & \textbf{Flare Max (UT)} & \textbf{Flare Class} & \textbf{EUV Wave Start (UT)} & \textbf{EUV Wave End (UT)} & \textbf{Source X ($"$)} & \textbf{Source Y ($"$)} & \textbf{CME on} & \textbf{$V_{CME}$} & \textbf{AW} \\
    \hline
    0 & 2010/06/12 & 0:30 & 0:57 & 20 & 0:55 & 1:19 & 633 & 390 & 1:32 & 486 & 119\\
    1 & 2010/08/14 & 9:38 & 10:05 & 4.4 & 9:30 & 10:08 & 697 & -26 & 10:12 & 1205 & 360\\
    2 & 2010/12/31 & 4:18 & 4:25 & 1.3 & 4:15 & 5:01 & 799 & 246 & 5:00 & 363 & 45\\
    3 & 2011/01/28 & 0:44 & 1:03 & 13 & 0:45 & 1:59 & 949 & 218 & 1:26 & 606 & 119\\
    4 & 2011/03/07 & 19:43 & 20:12 & 37 & 19:31 & 22:59 & 614 & 553 & 20:00 & 2125 & 360\\
    5 & 2011/05/11 & 2:23 & 2:43 & 0.81 & 2:20 & 2:35 & 785 & 399 & 2:48 & 745 & 225\\
    6 & 2011/08/04 & 3:41 & 3:57 & 93 & 3:43 & 4:20 & 546 & 200 & 4:12 & 1315 & 360\\
    7 & 2011/08/08 & 18:00 & 18:10 & 35 & 17:45 & 18:43 & 812 & 215 & 18:12 & 1343 & 237\\
    8 & 2012/03/07 & 1:05 & 1:14 & 130 & 0:00 & 0:40 & -475 & 397 & 1:30 & 1825 & 360\\
    9 & 2012/03/13 & 17:12 & 17:41 & 79 & 17:03 & 17:44 & 804 & 352 & 17:36 & 1884 & 360\\
    10 & 2012/07/23 & u & u & u & 2:09 & 2:48 & 912 & -243 & 2:36 & 2003 & 360\\
    11 & 2013/04/21 & u & u & u & 6:35 & 7:35 & 937 & 181 & 7:24 & 919 & 360\\
    12 & 2013/05/13 & 15:48 & 16:05 & 280 & 15:44 & 16:20 & -927 & 186 & 16:08 & 1850 & 360\\
    13 & 2013/05/15 & 1:25 & 1:48 & 120 & 1:06 & 1:50 & -852 & 199 & 1:48 & 1366 & 360\\
    14 & 2013/05/22 & 13:08 & 13:32 & 50 & 12:33 & 13:20 & 875 & 238 & 13:26 & 1466 & 360\\
    15 & 2013/06/21 & 2:30 & 3:14 & 29 & 2:31 & 3:21 & -869 & -268 & 3:12 & 1900 & 207\\
    16 & 2013/10/25 & 7:53 & 8:01 & 170 & 7:53 & 8:29 & -914 & -158 & 8:12 & 587 & 360\\
    17 & 2013/12/12 & 3:11 & 3:36 & 0.22 & 3:03 & 3:33 & 750 & -450 & 3:36 & 1002 & 276\\
    18 & 2013/12/28 & 17:53 & 18:02 & 9.3 & 17:10 & 18:00 & 942 & -252 & 17:36 & 1118 & 360\\
    19 & 2014/07/08 & 16:06 & 16:20 & 65 & 16:06 & 16:51 & -767 & 163 & 16:36 & 773 & 360\\
    20 & 2014/12/05 & 5:28 & 5:37 & 2.1 & 5:42 & 6:21 & 872 & -366 & 6:24 & 534 & 172\\
    21 & 2015/05/12 & 2:15 & 3:02 & 2.6 & 2:18 & 2:49 & 960 & -192 & 2:48 & 772 & 250\\
    22 & 2015/09/20 & 17:32 & 18:03 & 21 & 17:28 & 18:11 & 660 & -429 & 18:12 & 1239 & 360\\
    23 & 2015/10/29 & u & u & u & 2:13 & 2:52 & 951 & -167 & 2:36 & 530 & 202\\
    24 & 2015/11/09 & 12:49 & 13:12 & 39 & 12:51 & 13:27 & -626 & -229 & 13:25 & 1041 & 273\\
    25 & 2017/04/01 & 21:35 & 21:48 & 44 & 21:31 & 22:19 & 761 & 308 & 22:12 & 516 & 115\\
    \hline
\end{tabularx}
\end{table}

\begin{figure}[!htp] % updated!
  \centerline{\includegraphics[width=0.7\columnwidth]{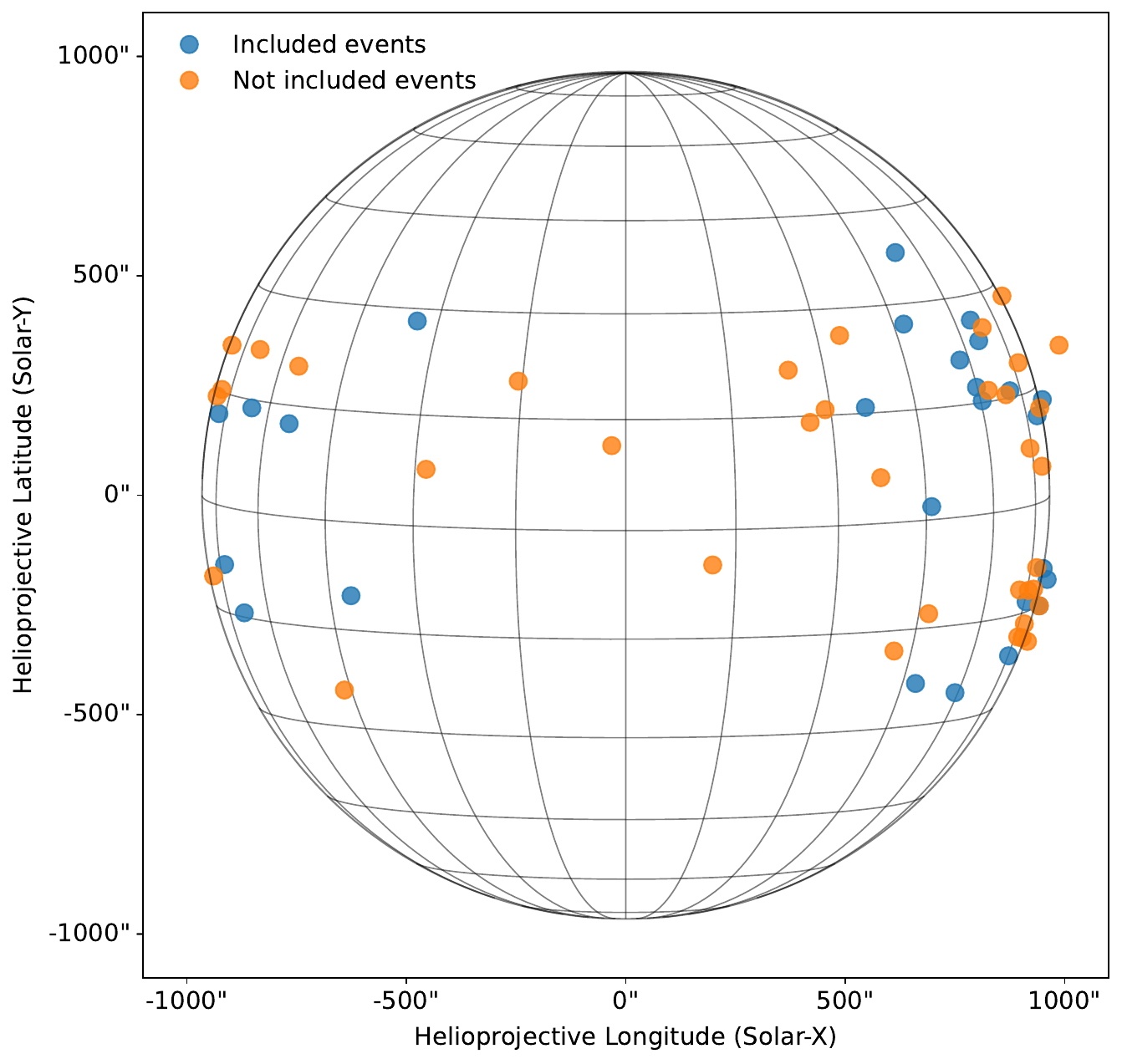}}
  \caption{Distribution of the source location of the CBFs on the solar disk. The blue dots are the events that we included in Table~\ref{table_1}, while the orange dots are the events that we did not include in the table.}
  \label{fig_solardisk}
\end{figure}

%Since all CBFs in our study were associated with flares, except for three events, we extracted their positions from the SOHO-LASCO Halo-CME catalog\footnote{LASCO CME Catalog: \url{https://cdaw.gsfc.nasa.gov/CME_list/halo/halo.html}}. This catalog provides the source locations of solar flares, which are assumed to be the eruption regions on the solar disk observed by the SOHO/EIT instrument \cite{delaboudiniere_1995}.
Figure~\ref{fig_solardisk} depicts the distribution of the CBFs on the Sun using the helioprojective coordinate system. The mean latitude and mean longitude of the CBFs were calculated as 56.35 and 378.04 arcsec, respectively. Additionally, the mean latitudes of CBFs in the northern and southern solar hemispheres were found to be 283.00 and -252.73 arcsec, respectively. As for the mean longitudes, they were -775.71 and 803.11 arcsec on the eastern and western sides, respectively.

CBFs appear relatively dim compared to the background solar disk. We found that channel 193 $\AA$ was most effective in clearly showing the wavefronts, although channel 211 $\AA$ proved better in some cases. Therefore, for each CBF event, we generated a sequence of base-difference images to study the evolution of CBFs. This involved subtracting the average of 10 images from all consecutive frames, with each frame separated by 24 seconds.

\begin{figure}[!htp] % updated!
  \centerline{\includegraphics[width=0.9\columnwidth]{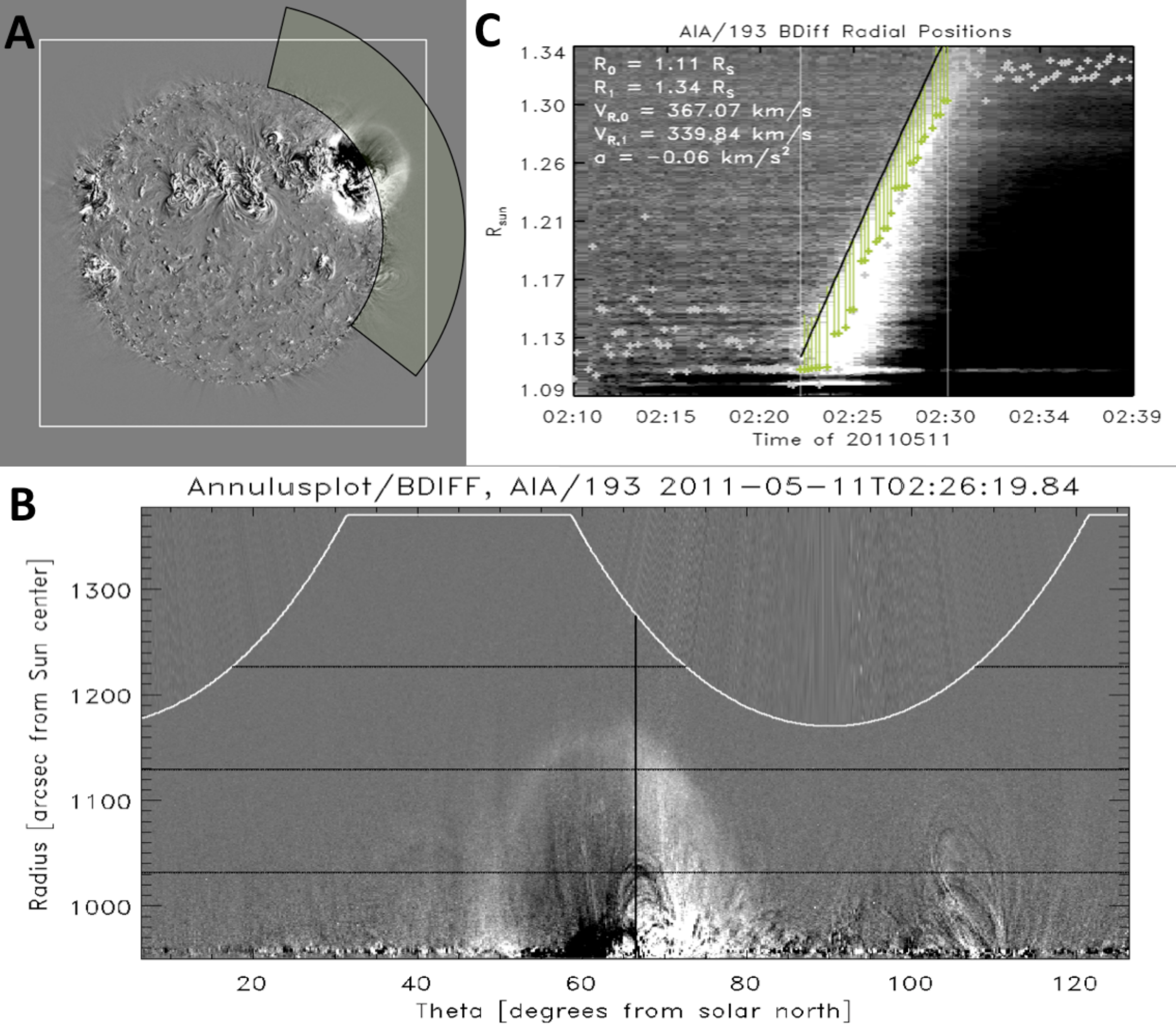}}
  \caption{Illustration for the annulus method used to extract kinematic data from AIA images. (A) shows the full Sun disk with the relevant region highlighted for analysis (green sector). The white box outlines the AIA FOV. (B) displays the extracted annular region mapped onto polar coordinates, with the actual data extent marked by the white curve. Black lines indicate the directions used for measuring radial and lateral motions. (C) shows a stacked plot of intensity along the radial direction, with green markers highlighting intensity peaks and their corresponding distances from the CBF wavefront. The white lines represent the time interval during which the CBF is tracked within the AIA FOV. This figure is curated from \cite{kozarev_2017}.}
  \label{fig_annplot}
\end{figure}

CBFs appear in AIA channels as quasi-spherical sheaths with brighter wavefronts, often interpreted as shock fronts \cite{vourlidas_2003, ontiveros_2009, kozarev_2011, ma_2011}.
To analyze their radial and lateral evolution, we applied the Coronal Analysis of SHocks and Waves framework \cite[CASHeW]{kozarev_2017}. This semi-automated technique involves extracting an annular region from AIA images and mapping it onto a polar projection (Fig.~\ref{fig_annplot}A). Intensity changes along radial and lateral directions are tracked to measure CBF kinematics. Users can interactively specify extraction lines and measure CBF positions (Fig.~\ref{fig_annplot}B,C). Extracted intensity pixels along the radial direction (CBF nose) throughout the event's duration within the AIA FOV are used to create a height-time plot (J-map).

Our analysis approximates radial and lateral wavefront positions using J-maps generated for each event. Assuming symmetrical expansion on both flanks, we treat the waves as spheroids defined by major and minor axes. Consequently, the radial direction is represented by a single value, while the lateral direction (parallel to the solar limb) uses two values for measurements in both left and right directions. However, lateral wave signatures may sometimes be visible in only one direction or be absent entirely.
Due to data limitations, our final sample size is 26 events, prioritizing complete datasets that exclude events with missing radial, lateral, or combined measurements.

To extract relevant plasma parameters and perform modeling, we retrieve information on CBFs from the HEK database and consult Nariaki Nitta's catalog of coronal waves  \cite{nitta_2013}\footnote{Nariaki Nitta's Catalog: \url{https://lmsal.com/nitta/movies/AIA_Waves/index.html}} to obtain necessary data. With the event list, we employ the SPREAdFAST framework to calculate kinematics, infer shock parameters, and determine plasma properties. Detailed summary plots of the modeled events can be found on the online SPREAdFAST catalog\footnote{SPREAdFAST Catalog: \url{https://spreadfast.astro.bas.bg/catalog/}}.

\begin{figure}[!htp] % updated!
  \centerline{
      \includegraphics[width=0.32\textwidth,clip=]{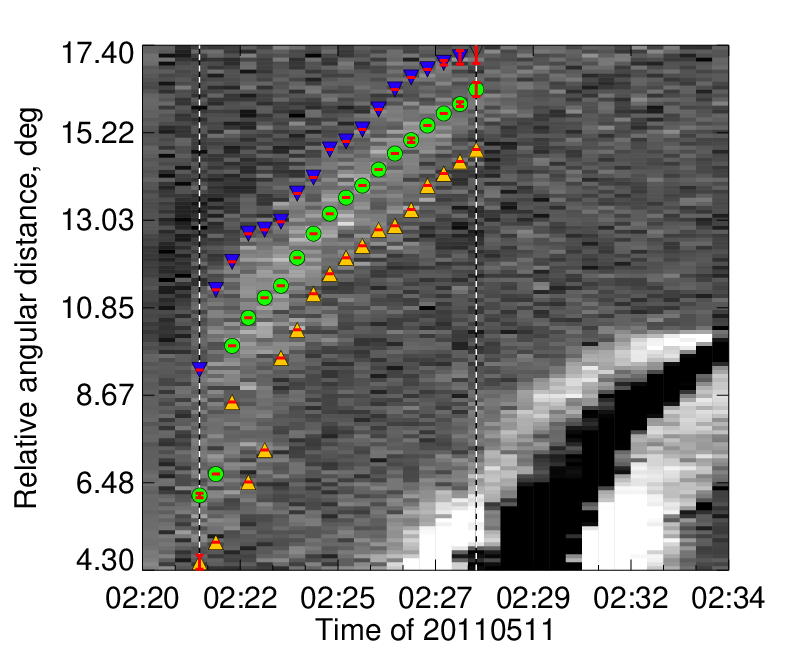}
      \includegraphics[width=0.32\textwidth,clip=]{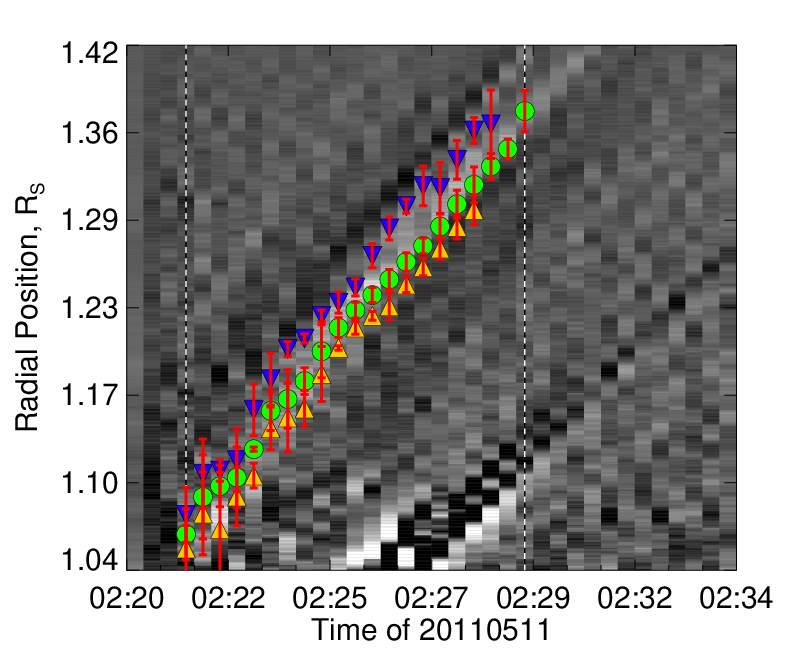}
      \includegraphics[width=0.32\textwidth,clip=]{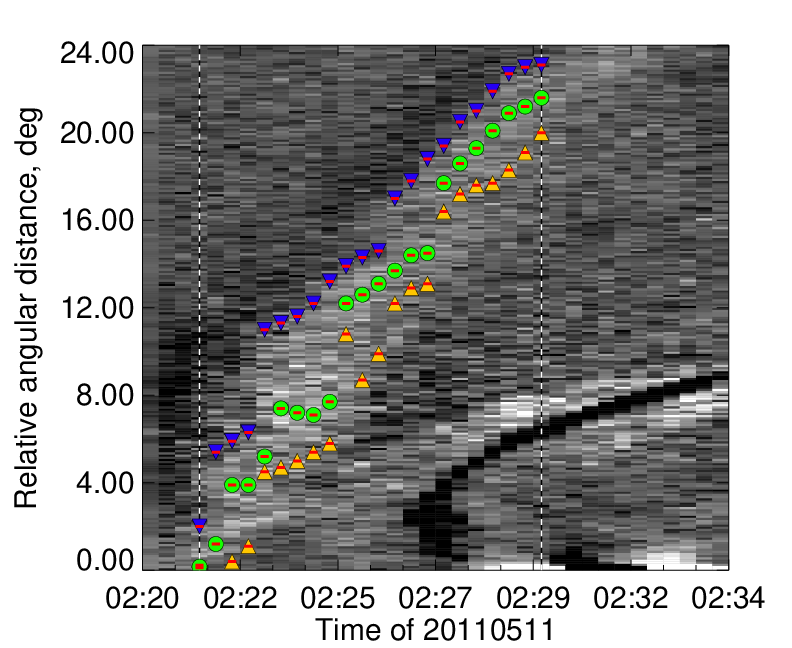}
      }
  \caption{J-map plots for the event of May 11, 2011, for the radial direction (middle plot) and the left and right flanks of the wave in the lateral heliocentric direction (the left and right plots, respectively). Blue, green, and orange filled symbols are the positions of the CBF front, peak, and back, respectively. The uncertainty of the average measurements is shown as red bars.}
  \label{fig_jmaps_110511}
\end{figure}

To accurately determine the front, back, and peak of the EUV wave at each time step (Fig.~\ref{fig_jmaps_110511}), we applied several algorithms. Firstly, we utilized Savitzky-Golay filtering \cite{savitzky_1964} to smoothen the data. Next, we employed local minima/maxima ordering and proximity/intensity metrics algorithms. These algorithms enabled us to identify the wave positions and extract relevant parameters.
For each CBF event, we manually specified the starting and ending times, indicated by vertical white lines in Figure~\ref{fig_jmaps_110511}. We also determined the starting and ending height, corresponding to the off-limb portion of the CBF within the AIA FOV.

By analyzing the intensity values, we defined the CBF positions as the locations with peak intensity at each time step. The front and back of the wave were set at 20\% of the peak intensity.
To obtain more comprehensive information about the CBFs, we applied the Levenberg-Marquardt least squares minimization \cite{markwardt_2009} along with a second-degree bootstrapping optimization technique \cite{efron_1979}. This approach allowed us to fit fourth-order polynomials to the wave positions using Savitzky-Golay fitting. As a result, we obtained measurements for speeds, accelerations, intensities, and thicknesses of the waves in both the radial and lateral directions. The thicknesses and intensities are averages of the values between the peak and back for each time step. Measurements of the heights of CBFs with respect to the solar disk center were obtained\footnote{LASCO CME Catalog: \url{https://cdaw.gsfc.nasa.gov/CME_list/}} for all the events analyzed in this study. These measurements were taken at the fastest segment of the leading edge of each CBFs over time.
We also have measurements of the lateral positions of the CBF front relative to the nose direction. These were obtained in degrees and converted to km depending on the height of the lateral measurement above the solar surface.

\subsection{Analysis Methods}
The Solar Particle Radiation Environment Analysis and Forecasting–Acceleration and Scattering Transport \cite[SPREAdFAST]{kozarev_2022} is a physics-based prototype heliospheric SEP forecasting system. It incorporates data-driven models to estimate the coronal magnetic field, dynamics of coronal shock waves, energetic particle acceleration, and scatter-based SEP propagation in the heliosphere. The system is based on the CASHeW framework \cite{kozarev_2017} and provides timely predictions of SEP arrival times, maximum intensities, and SEP fluxes at various locations in the inner heliosphere. It contributes to space weather requirements, protecting ESA assets, aiding satellite operators, and providing lead times for mitigating impacts on electronics and humans in space activities.

Summary plots of the J-maps, including estimated positions and errors, can be found in the online SPREAdFAST catalog for each event. To create a unified lateral kinematics time series for each event, we average measurements from both lateral flanks. Additionally, we record the CBF mean intensity and thickness in both directions.
To analyze the kinematic measurements deduced from the AIA FOV, we apply a Savitzky-Golay fit \cite{savitzky_1964}, as described in Kozarev et al.\cite{kozarev_2019}. Subsequently, we extrapolate the smoothed radial positions up to $\sim$17\rsun using the analytical CME kinematics models presented by Gallagher et al.\cite{gallagher_2003} and Byrne et al.\cite{byrne_2013}.

Our next step involves developing multiple synthetic geometric shock models, known as the synthetic shock model (S2M) module, to describe the shock surface at a 24-second cadence. These models rely on extrapolated radial and lateral kinematic results, as well as the inferred major and minor axes of the spheroids representing compressive waves. The shock surface is created from the onset of the CBF until its nose reaches 10\rsun and is then propagated up to $\sim$17\rsun. The propagation is based on the Magnetohydrodynamic Algorithm outside a Sphere (MAS) synoptic coronal MHD model's results. Consequently, the shock surface samples plasma parameters from the data cube of the MAS model at discrete points, determined by consecutively crossing magnetic field lines.
The MHD data utilized in this study is represented as a 3D data cube consisting of plasma parameters. To analyze this data, a spheroid model was propagated through the data cube by scanning it without any direct interaction. At each point within the data cube, a search was conducted to identify the nearest four neighbors. By employing trilinear interpolation, the values at these points were estimated.

By sampling the shock surface, we obtain data for approximately 1000 field-crossing lines, potentially more depending on the desired resolution. For each event, the output consists of a set of data structures that describe each shock-crossing field line. These structures include the shock speed $V_{shock}$, plasma density $n$, density jump $r$, shock upstream magnetic field magnitude $B_{mag}$, shock-field angle $\theta_{BN}$, Alfven speed $V_A$, and Alfven Mach number $M_A$.

\begin{figure}[!htp] % updated!
  \centerline{\includegraphics[width=0.6\columnwidth]{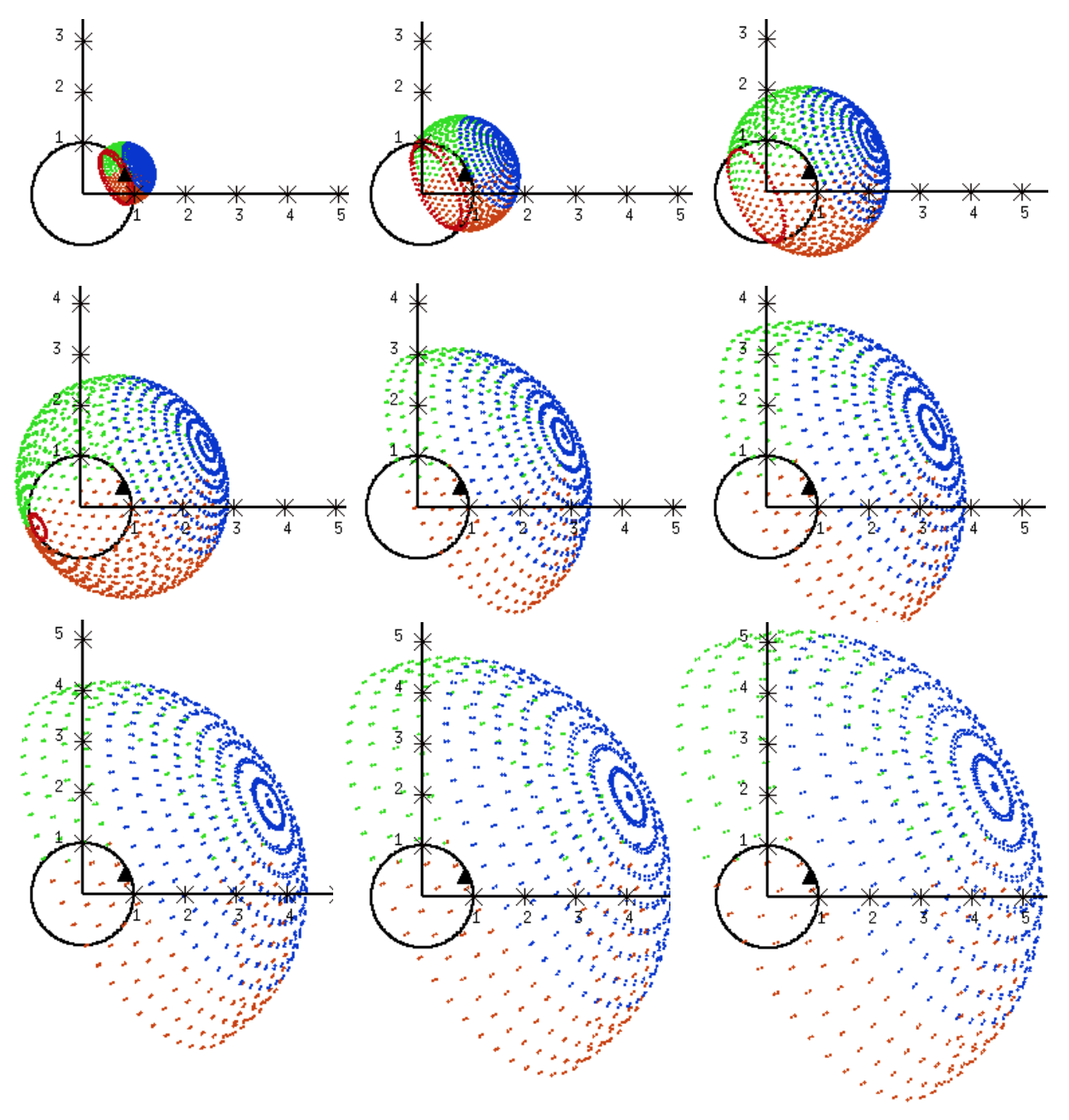}}
  \caption{Synthetic shock model divided into three segments; the cap zone in blue and the flank zones are in red and green.}
  \label{fig_segments}
\end{figure}

To estimate the shock density jump, we follow the method of \cite{kozarev_2017} - we calculate the differential emission measure (DEM) before and during the event at each shock crossing and each timestep. The DEM is obtained using the model by \cite{cheung_2015}. We integrate the DEM to obtain the average density, and take the ratio of densities during and before the event.
Typically, the density jump within the AIA FOV is relatively small, usually below 1.2. However, beyond this region where observational information is lacking, we assign a value of 1.2, assuming a weak shock.
For ease of analysis, we divide the synthetic shock model into three segments: the cap (representing the shock nose), Zone 1, and Zone 2 (referring to the shock flanks). This division allows us to examine the distribution of plasma parameters across different sectors of the shock surface. Figure~\ref{fig_segments} illustrates the evolution of the synthetic shock model in nine timesteps, with the cap zone colored blue and the shock flanks colored green and red.

%%________________________________________________________________
\section{Case Study: May 11, 2011}
\label{s_case_study}
In this section, we provide a detailed analysis of the event at the low corona region, demonstrating our method. Additionally, we investigate the plasma parameters along individual shock-crossing magnetic field lines in the AIA FOV.

\subsection{Event Context}
\begin{figure}[!htp] % updated!
  \centerline{\includegraphics[width=1\columnwidth]{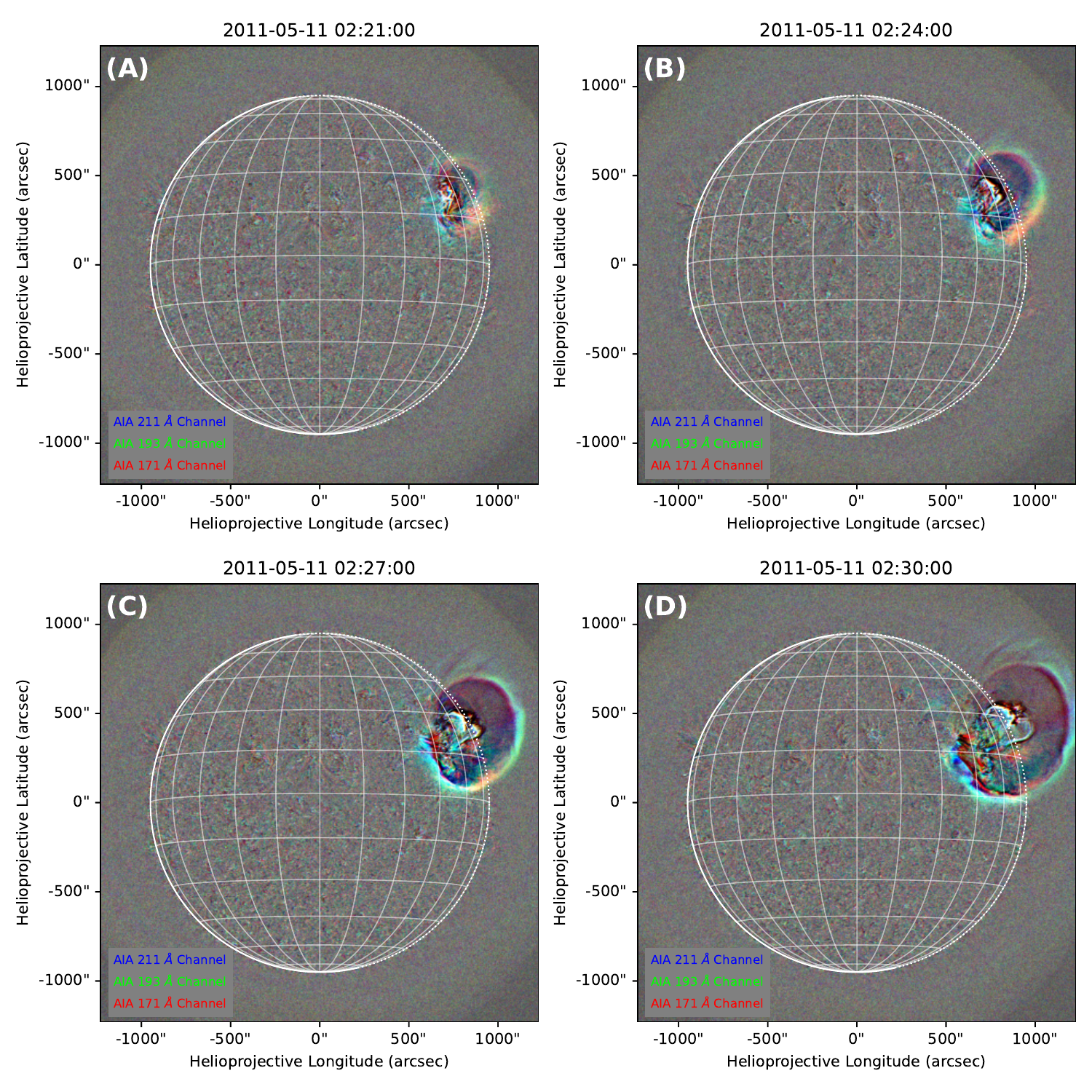}}
  \caption{AIA running-difference images capture a coronal wave evolving over 9 minutes near the Sun's western limb, exhibiting markedly changing intensity and structure as observed in 171, 193, and 211 \AA.}
  \label{fig_aia_event}
\end{figure}

The eruption took place on May 11, 2011, at approximately 02:20 UT (Fig.~\ref{fig_aia_event}). It originated from an active region situated in the northwestern sector (N18W52). The event involved a massive shock wave propelled by a fast partial-halo CME that occurred at 02:48 UT. The CME exhibited a linear speed of 745 \kms, a 2$^{nd}$-order speed at 20\rsun of 776 \kms, an acceleration of 3.3 m s$^{-2}$, an angular width (AW) of 225\degree, a central position angle (PA) of 320\degree, a measurement position angle (MPA) of 283\degree, a mass of 3.5$\times$10$^{15}$ gram, and a kinetic energy of 9.6$\times$10$^{30}$ erg. The mass and kinetic energy were uncertain due to projection effects, as reported by the SOHO-LASCO CME catalog. This was accompanied by a weak solar flare classified as B8.1 and an eruptive filament, as observed by the 193 $\AA$ EUV channel of the SDO/AIA.

Furthermore, the eruption was associated with a type II radio burst, which commenced around 02:20 UT. This was observed by the Learmonth spectrogram (25-180 MHz) maintained by the Australian Space Weather Services and part of the CALLISTO global network. By examining the OMNI database\footnote{OMNIWeb Database: \url{https://omniweb.gsfc.nasa.gov/}}, we found no evidence of a geomagnetic storm occurring within three days from the onset of the eruption. Nevertheless, an increase in proton fluxes across all energy channels near 1 AU was observed using the SOHO/ERNE instrument. According to the Wind/EPACT catalog\footnote{Wind/EPACT Catalog: \url{http://newserver.stil.bas.bg/SEPcatalog/}} \cite{miteva_2016, miteva_2017}, we found an SEP event detected by the SOHO/ERNE instrument at the Earth with onset time of 03:39 UT and a $J_p$ of 0.0133 protons/(cm$^2$ s sr MeV) in the energy channel 17-22 MeV. $J_p$ is the peak proton intensity after subtracting the pre-event level.

\subsection{Low Corona Part}
To investigate the kinematics of the CBF event, we employed the CASHeW module within the SPREAdFAST framework. As we see in Figure~\ref{fig_jmaps_110511}, the J-maps are displayed, illustrating the radial and lateral time-dependent evolution of the CBF in gray-scale. Since the wave is assumed to have a dome-like shape, the lateral direction is divided into left and right flanks. Bright features below the CBF in the J-maps are likely expanding loops. To estimate the uncertainty in the measurements, we varied slightly the radial (nose) direction three times. The corresponding positions are depicted in red, while the start and end times of the CBF are indicated by vertical dashed lines. Additionally, the front, back, and peak of the CBF are represented by blue down-pointing triangles, yellow up-pointing triangles, and green-filled circles, respectively.

\begin{figure}[!htp] % updated!
  \centerline{
      \includegraphics[width=1\textwidth,clip=]{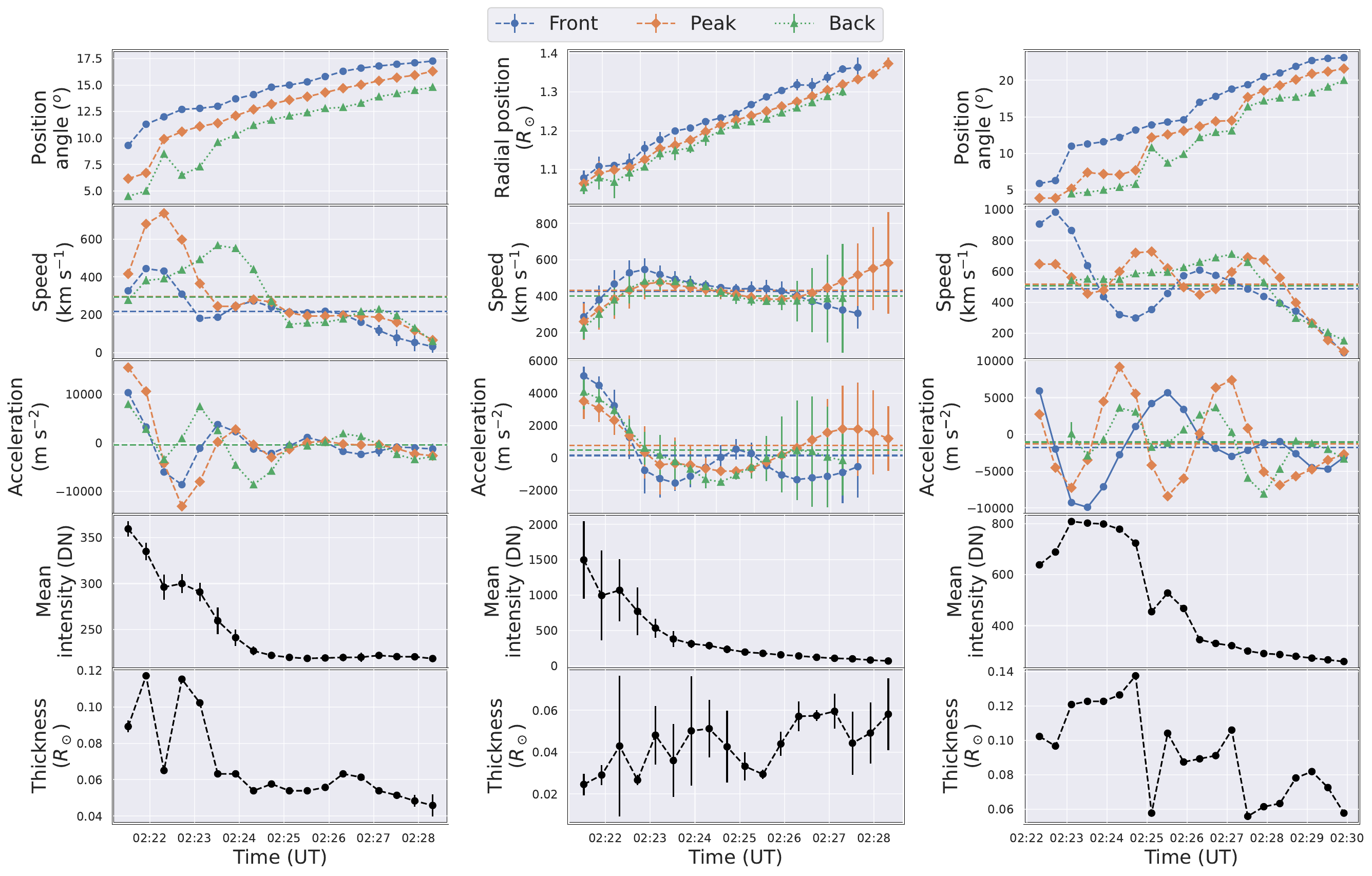}
      }
  \caption{Time-series kinematics of the CBF parameters for the front, peak, and back positions in the AIA FOV, with measurement uncertainties shown as small bars over the data points. The horizontal lines in the speed and acceleration panels denote the mean speeds and accelerations for the wave front, peak, and back with respective colors. The left and right columns represent the lateral kinematic measurements in the left and right flanks of the wave, respectively. The middle column represent the kinematic measurements in the radial direction.}
  \label{fig_kinematics_110511}
\end{figure}

Figure~\ref{fig_kinematics_110511} presents the time series kinematic results of the shock wave parameters within the SDO/AIA field of view (up to 1.3\rsun). The kinetics of the wavefront, peak, and back are color-coded as red, green, and blue, respectively. The subpanels from top to bottom display the estimated heliocentric distance, speed, acceleration, intensity, and thickness of the wave. These parameters are presented for both the radial (middle panel) and lateral directions (left and right panels).

Analysis of Figures~\ref{fig_jmaps_110511} and~\ref{fig_kinematics_110511} reveals that the coronal wave was asymmetric in shape. The time-dependent evolution of the angular distance differed slightly between the left and right flanks. In Figure~\ref{fig_jmaps_110511}, the right flank (towards the solar equator) of the wave appeared for a little bit longer time, allowing the algorithm to capture it with a higher number of measurements until approximately 02:29 UT, same as for the radial direction. In contrast, the left flank (towards the solar pole) had fewer measurements available.

The coronal wave's initial appearance was slightly elongated, with an aspect ratio of 0.5. This indicates a longer major axis, creating a degree of asymmetry. At 02:25:31 UT, a striking change occurred: the wave became perfectly circular, achieving an aspect ratio of 1. This signifies equal lengths for both axes, resulting in a symmetrical shape. However, this transformation was short-lived. The wave's morphology shifted again, becoming increasingly flattened. This signifies a growing minor axis compared to the major one, leading to an \textit{over-expansion} of the wave along its minor axis.
In this study, the aspect ratio is defined as the minor axis divided by the major axis of the wave's geometric surface. A value of 1 represents a perfectly symmetrical wave, while values greater than 1 indicate over-expansion along the minor axis. Conversely, values less than 1 point towards elongation along the major axis, reflecting a more radial expansion.

Regarding the radial direction, the event duration spanned from approximately 02:21 to 02:28 UT. The shock wave exhibited an average speed of approximately 420.46 \kms, while the average acceleration was around 463.92 m s$^{-2}$, calculated as the mean of the front, peak, and back sides of the wave. For the left flank in the lateral direction, the event duration spanned from around 02:21 to 02:28 UT. The average speed and acceleration were approximately 270 \kms and -400.62 m s$^{-2}$, respectively. For the right flank in the lateral direction, the event duration spanned from around 02:21 to 02:30 UT, lasting for one minute longer than the left flank. The average speed and acceleration were approximately 500.97 \kms and -297.18 m s$^{-2}$, respectively.

Comparing the lateral directions, the wave's sheath on the right flank was approximately six times the thickness observed on the left flank, while the radial direction exhibited a thickness roughly half that of the left flank. Notably, the peak speed in the radial direction was lower than that in the lateral direction; right flank, suggesting that the shock wave experienced compression in the direction of propagation, while expanding laterally to a greater extent than radially. Table~\ref{T_110511} provides a summary of the statistical results.

To further explore the shock and plasma parameters at different sections of the coronal wave, we divided the shock surface into three segments: the Cap zone (shock nose), Zone 1, and Zone 2 (the shock flanks). This division is illustrated in Figure~\ref{fig_segments}.

\begin{table}[!htp] % updated!
\centering
\caption{Mean values and their standard deviation of the wave parameters in the radial direction and the lateral direction for the left and right flanks, at the front, peak, and back sides of the wave for the event occurred on May 11, 2011, in the SDO/AIA FOV.}
\label{T_110511}
\resizebox{\textwidth}{!}{%
\begin{tabular}{lc|c|c|c}
\hline
Parameter                              & Direction  & Front              & Peak                   & Back                 \\ \hline
\multirow{3}{*}{$<speed>$ \kms}        & Lat. Left  & 218.46 $\pm$ 9.04  & 297.46 $\pm$ 5.45      & 293.94 $\pm$ 9.04    \\ \cline{2-5}
                                       & Radial     & 427.46 $\pm$ 51.85 & 433.11 $\pm$ 82.86     & 400.81 $\pm$ 83.78   \\ \cline{2-5} 
                                       & Lat. Right & 494.69 $\pm$ 0.00  & 509.25 $\pm$ 1.02      & 498.97 $\pm$ 9.21    \\ \hline
\multirow{3}{*}{$<accel.>$ m s$^{-2}$} & Lat. Left  & -414.62 $\pm$ 227.23 & -401.46 $\pm$ 164.62 & -385.77 $\pm$ 227.23 \\ \cline{2-5}
                                       & Radial     & 147.41 $\pm$ 1009.19 & 758.97 $\pm$ 1287.65 & 485.38 $\pm$ 1365.80 \\ \cline{2-5} 
                                       & Lat. Right & -415.04 $\pm$ 0.00   & -209.81 $\pm$ 22.32  & -266.68 $\pm$ 250.80 \\ \hline
\multirow{3}{*}{$<intensity>$ DN}      & Lat. Left  & \multicolumn{3}{c}{250.60 $\pm$ 5.90}               \\ \cline{2-5}
                                       & Radial     & \multicolumn{3}{c}{403.34 $\pm$ 143.30}             \\ \cline{2-5}
                                       & Lat. Right & \multicolumn{3}{c}{489.04 $\pm$ 2.86}               \\ \hline
\multirow{3}{*}{$<thickness>$\rsun}   & Lat. Left  & \multicolumn{3}{c}{0.07 $\pm$ 0.00}                 \\ \cline{2-5}
                                       & Radial     & \multicolumn{3}{c}{0.04 $\pm$ 0.01}                 \\ \cline{2-5}
                                       & Lat. Right & \multicolumn{3}{c}{0.09 $\pm$ 0.00}                 \\ \hline
\end{tabular}%
}
\end{table}

% \kkozarev{Theta$_{BN}$ can't be thousands of degrees - fix by subtracting iteratively 360\degree degrees until it is between 0-90\degree.} \mnedal{I tried to subtract/add 360\degree iteratively and didn't work. I had to subtract/add 90\degree instead until the angle is between 0-90\degree.}
\begin{table}[!htp] % updated!
  \centering
  \caption{Mean, median, and standard deviation of the shock parameters output, from the interaction of the S2M spheroid with the MAS MHD model results, for the shock's cap and flanks and for the whole shock surface, for the event on May 11, 2011.}
  \label{T_sh_param_110511}
  \begin{tabular}{lcccc}
    \hline
    \multirow{2}{*}{Segment} & \multirow{2}{*}{Parameter} & \multicolumn{3}{c}{Statistics} \\
    &                         & Mean & Median & Stdv \\ \hline
    All & $V_{SHOCK}$ \kms    & 577.77 & 578.39 & 72.79 \\ 
    & $\theta_{BN}$ \degree   & 70.06 & 0.63 & 44.83 \\ 
    & $B_{MAG}$ G             & 0.046 & 0.038 & 0.070 \\ 
    & Density Jump            & 1.193 & 1.188 & 0.185 \\ \hline
    
    Cap & $V_{SHOCK}$ \kms    & 555.18 & 550.86 & 42.46 \\ 
    & $\theta_{BN}$ \degree   & 19.37 & 3.61 & 25.51 \\ 
    & $B_{MAG}$ G             & 0.046 & 0.036 & 0.070 \\ 
    & Density Jump            & 1.193 & 1.188 & 0.015 \\ \hline
    
    Zone 1 & $V_{SHOCK}$ \kms & 613.69 & 609.32 & 59.42 \\ 
    & $\theta_{BN}$ \degree   & 6.46 & 0.21 & 50.92 \\ 
    & $B_{MAG}$ G             & 0.045 & 0.045 & 0.066 \\ 
    & Density Jump            & 1.190 & 1.187 & 0.008 \\ \hline
    
    Zone 2 & $V_{SHOCK}$ \kms & 631.37 & 614.23 & 73.07 \\ 
    & $\theta_{BN}$ \degree   & 0.10 & 0.51 & 10.61 \\ 
    & $B_{MAG}$ G             & 0.046 & 0.029 & 0.071 \\ 
    & Density Jump            & 1.194 & 1.188 & 0.016 \\ \hline
  \end{tabular}
\end{table}
  
We summarize the results for the three segments in Table~\ref{T_sh_param_110511} to further investigate the shock and plasma parameters in different sections of the coronal wave. Notably, the mean shock speed at the flanks was higher than that at the Cap zone. We did not observe significant variations in the magnetic field across the different segments, indicating a relatively homogeneous magnetic structure. The shock density jump exhibited consistent values across all three segments.

In \cite{kozarev_2022} we investigated shock-crossing magnetic field lines during this event, and key plasma parameters were analyzed up to 10\rsun. The study, utilizing DEM analysis, revealed consistent results with weak coronal shocks. Notably, the density jump within the AIA FOV was generally small, below 1.2, aligning with previous research. Beyond this view, lacking observational data, the density jump was set to 1.2.
By inspecting the parameter evolution over all shock-crossing field lines, we found that the shock-field angle ($\theta_{BN}$) and magnetic field amplitude ($|B|$) consistently decreased over time and radial distance.

The crucial parameter for diffusive shock acceleration (DSA), $\theta_{BN}$, was further detailed, highlighting its time-dependent distribution across the entire spheroid surface. Notably, there was a significant decrease in $\theta_{BN}$ angle within the first 50 minutes of the event. Additionally, dividing the spheroid into distinct regions revealed nuanced variations, with the cap/nose region exhibiting the lowest $\theta_{BN}$ values, while zone 2 consistently displayed higher values above 60\degree. All dynamic spectra for individual events are accessible on the SPREAdFAST catalog webpage.

\subsection{Middle/Outer Corona Part}
We collected complementary measurements from the SOHO/LASCO instrument in order to expand the analysis of EUV waves' kinematics in the middle/outer corona. These measurements specifically provide the radial distance of the CME leading edge associated with the coronal wave over time, which is referred to as the height-time profile of the CME.

\begin{figure}[!htp] % updated!
  \centerline{\includegraphics[width=0.9\columnwidth]{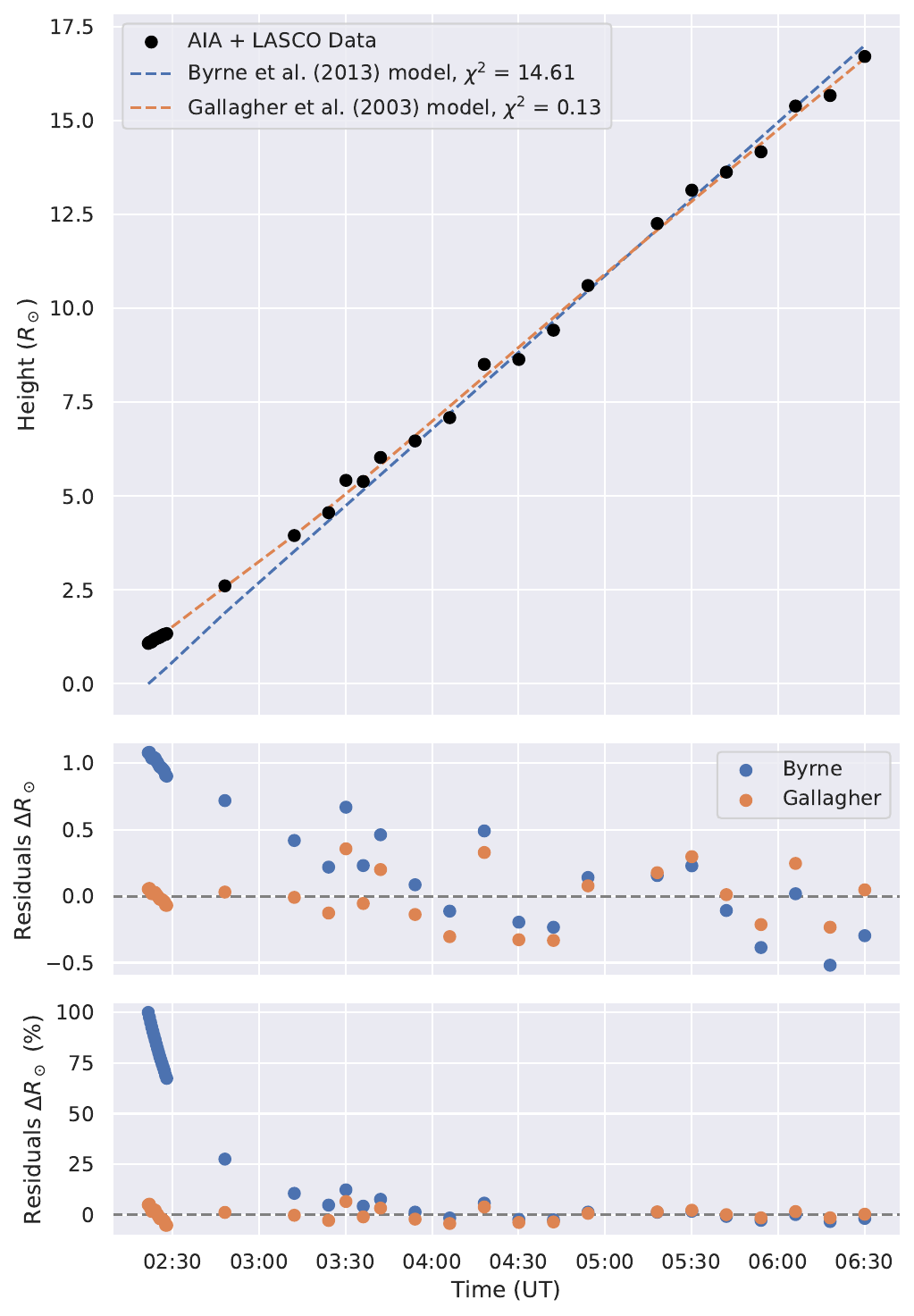}}
  \caption{Top panel -- Height-time profile compiled from AIA and LASCO measurements for the event occurred on May 11, 2011, fitted with two CME kinematics models from the photosphere up to 17\rsun. Middle panel -- Difference between the fitting and the real observations. Bottom panel -- Relative residuals in \%.}
  \label{fig_height_profile_aialasco_110511}
\end{figure}

Figure~\ref{fig_height_profile_aialasco_110511} displays the extended measurements of the EUV in the LASCO/SOHO FOV, reaching approximately 17\rsun. To analyze the height measurements, we applied the models of CME kinematics proposed by \cite{byrne_2013} and \cite{gallagher_2003}. Through a comparison, we determined that the model by \cite{gallagher_2003} provided a better fit with a $\chi^2$ value of 0.13. Examining the bottom panel of the figure, we observe the residuals (i.e. the differences between the actual measurements and the fits) for both models. It becomes apparent that the residuals are generally lower for the Gallagher model when compared to the Byrne model.
The efficacy of the Gallagher fitting model in accommodating both AIA and LASCO measurements is underscored by Figure~\ref{fig_height_profile_aialasco_110511}. This alignment underscores the model's proficiency in capturing the early stages of the solar event, particularly near the Sun.

\begin{figure}[!htp] % updated!
  \centerline{\includegraphics[width=0.9\columnwidth]{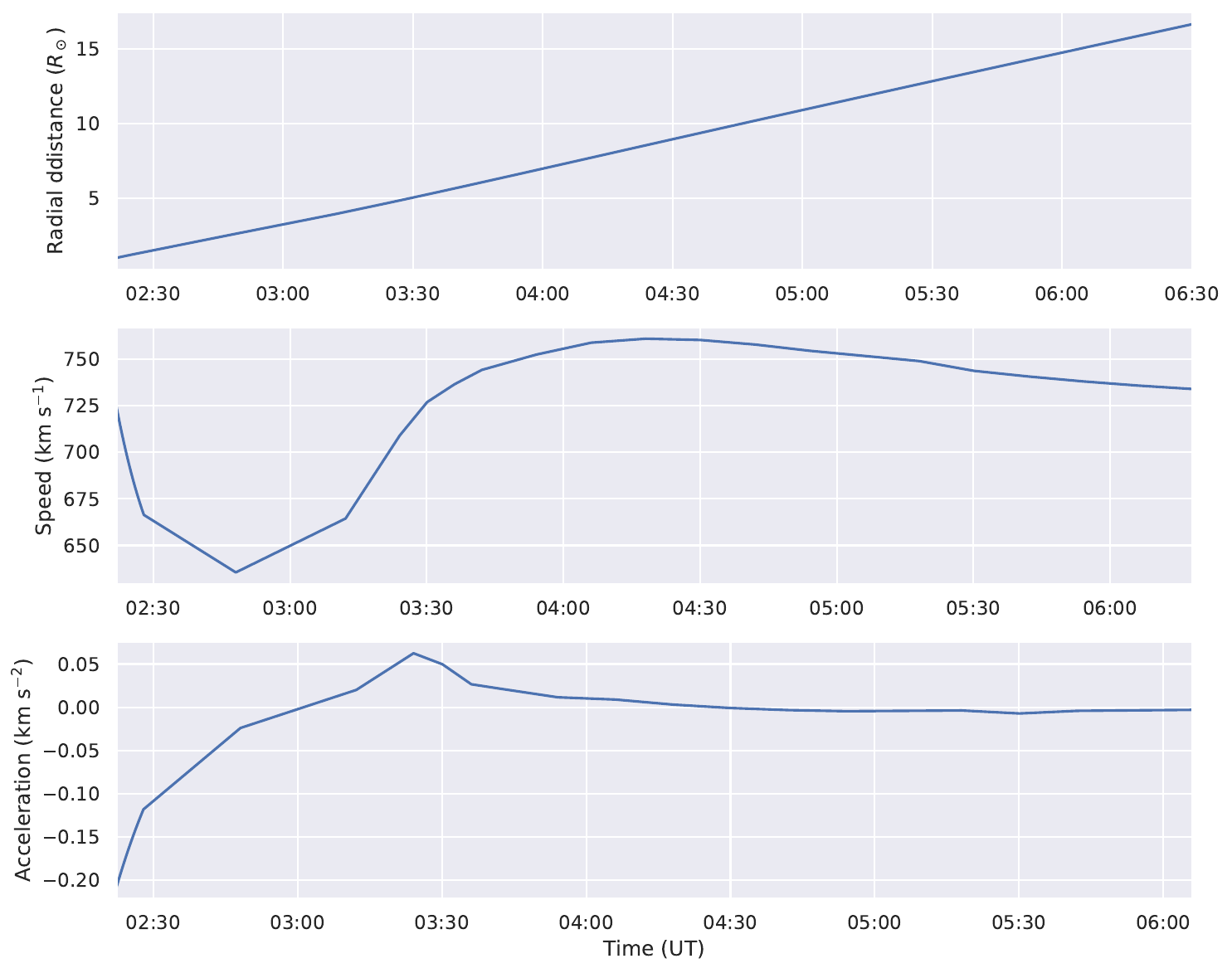}}
  \caption{Extrapolated radial kinematics for the event occurred on May 11, 2011, based on the ballistic model of \cite{gallagher_2003} up to 17\rsun.}
  \label{fig_rad_kinematics_aialasco_110511}
\end{figure}

Further insights can be gained by examining Figure~\ref{fig_rad_kinematics_aialasco_110511}, which demonstrates that the wave experienced a period of rapid acceleration between approximately 02:25 and 03:15 UT within a distance of approximately 5\rsun from the Sun. This behavior aligns with the fluctuations in wave acceleration depicted in Figure~\ref{fig_kinematics_110511}. At the same time, the wave speed had a sharp decrease from around 727 \kms to 570 \kms within an hour. Subsequently, the wave speed gradually increased over the following 3 hours, covering a distance of around 15\rsun and they it plateaued at approximately 723 \kms.

%______________________________________________________________
\section{Statistical Study}
\label{s_stat_study}
We present a comprehensive statistical analysis of the kinematic characteristics and plasma parameters of coronal wave events observed in the AIA and LASCO FOVs.

An overview of the statistical parameters related to shock characteristics, including wave speed, intensity, and thickness in the AIA FOV, is presented in Table~\ref{T_sh_param_all}. The wave speeds are expressed in \kms, wave accelerations are in km s$^{-2}$, wave intensity in arbitrary units, and wave thickness in\rsun, as the data have undergone multiple stages of processing.

Upon analyzing the data, we observed that the waves generally exhibited higher speeds, higher acceleration, lower mean intensities, and lower thickness in the radial direction compared to the lateral direction. This suggests that the waves were somewhat elongated in their early stages near the Sun, potentially due to the coronal conditions, including plasma densities and magnetic field strength and structure.

\begin{figure}[!htp] % updated!
  \centerline{
      \includegraphics[width=0.32\textwidth,clip=]{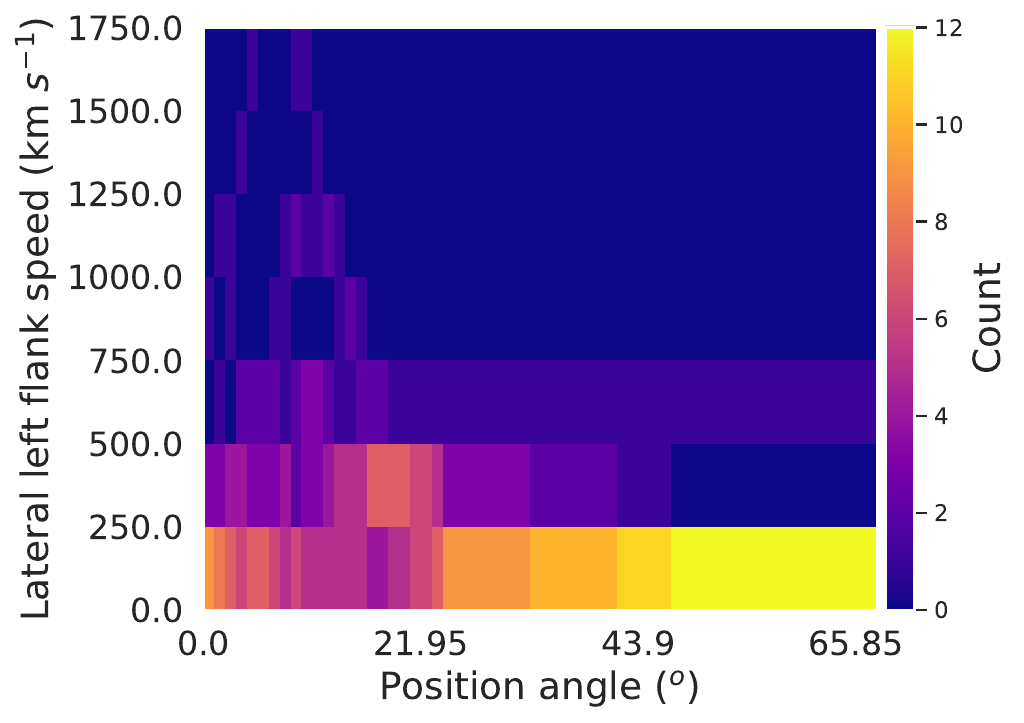}
      \includegraphics[width=0.32\textwidth,clip=]{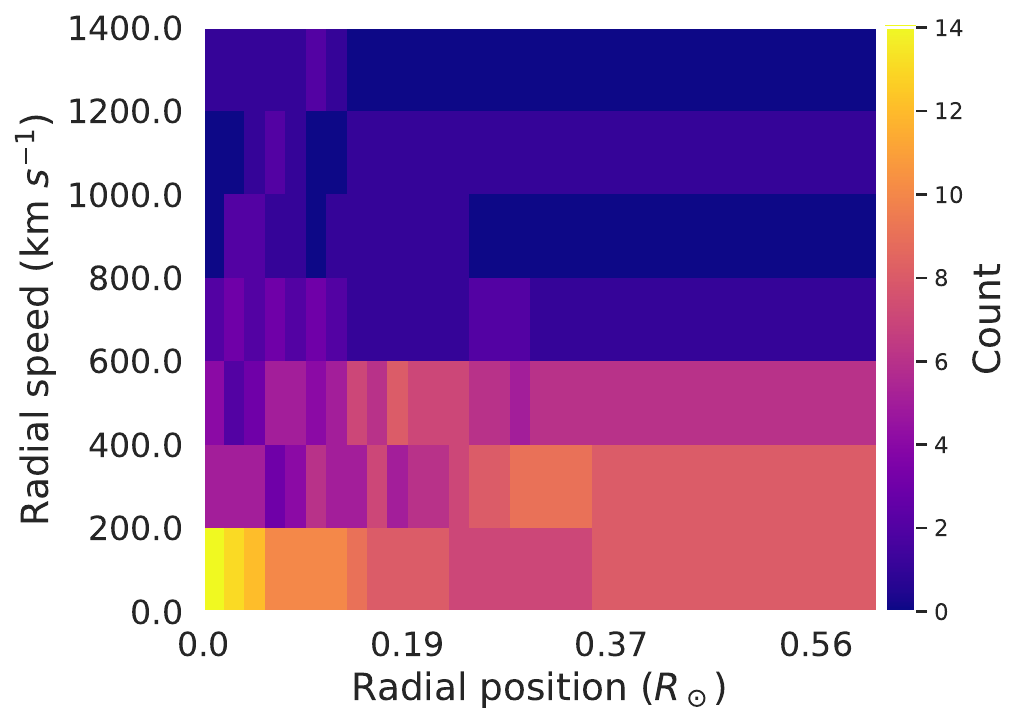}
      \includegraphics[width=0.32\textwidth,clip=]{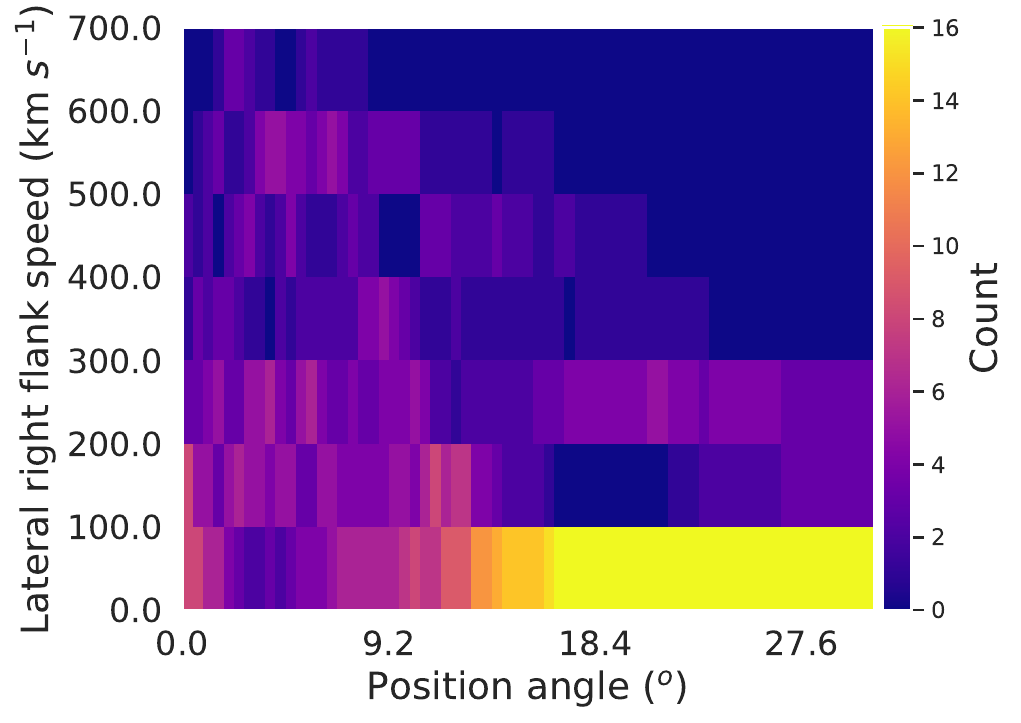}
      }
  \centerline{
      \includegraphics[width=0.32\textwidth,clip=]{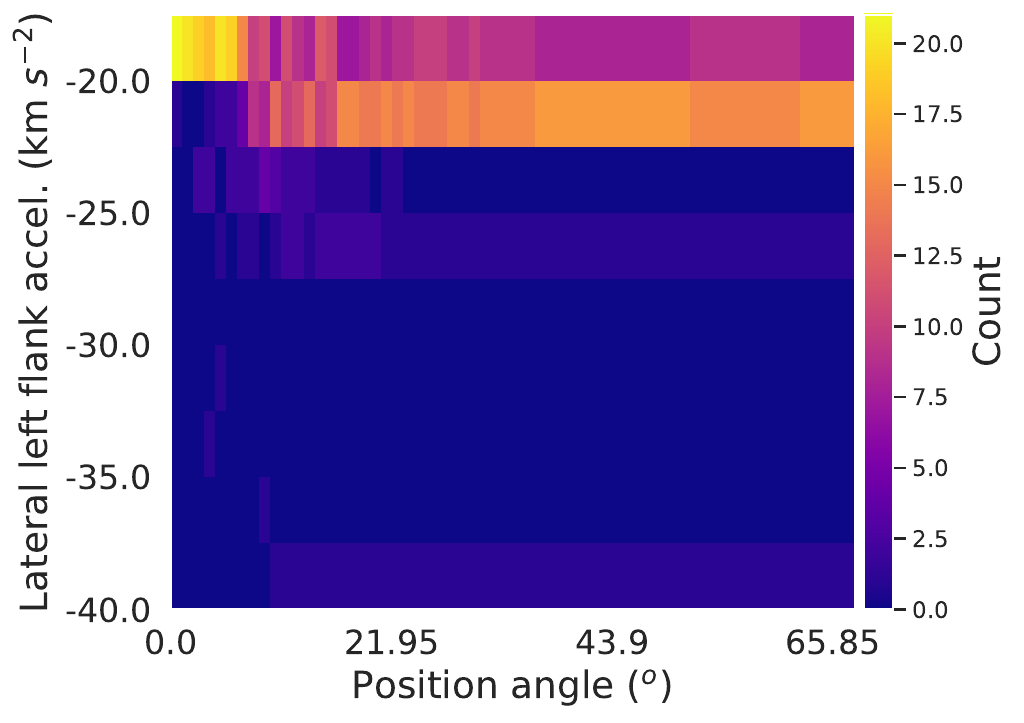}
      \includegraphics[width=0.32\textwidth,clip=]{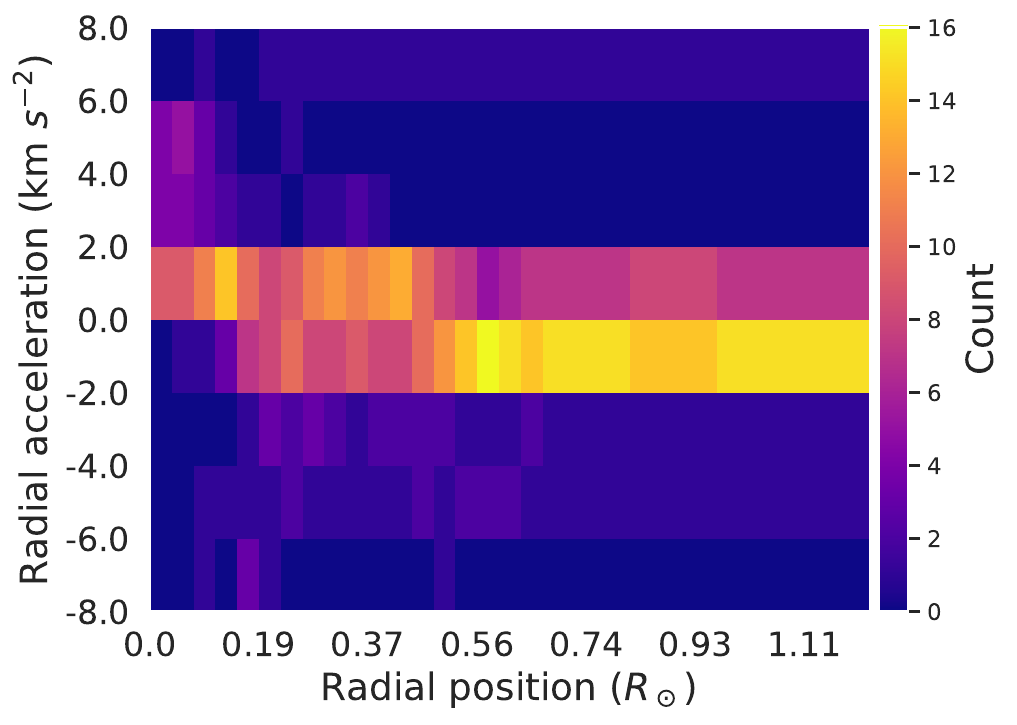}
      \includegraphics[width=0.32\textwidth,clip=]{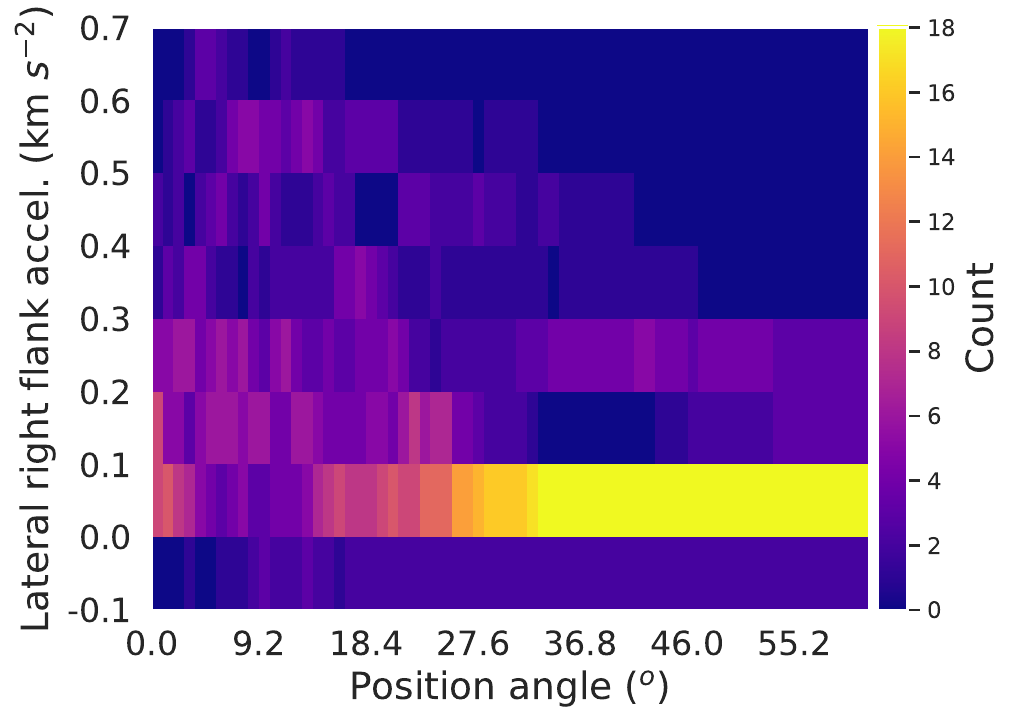}
      }
  \centerline{
      \includegraphics[width=0.32\textwidth,clip=]{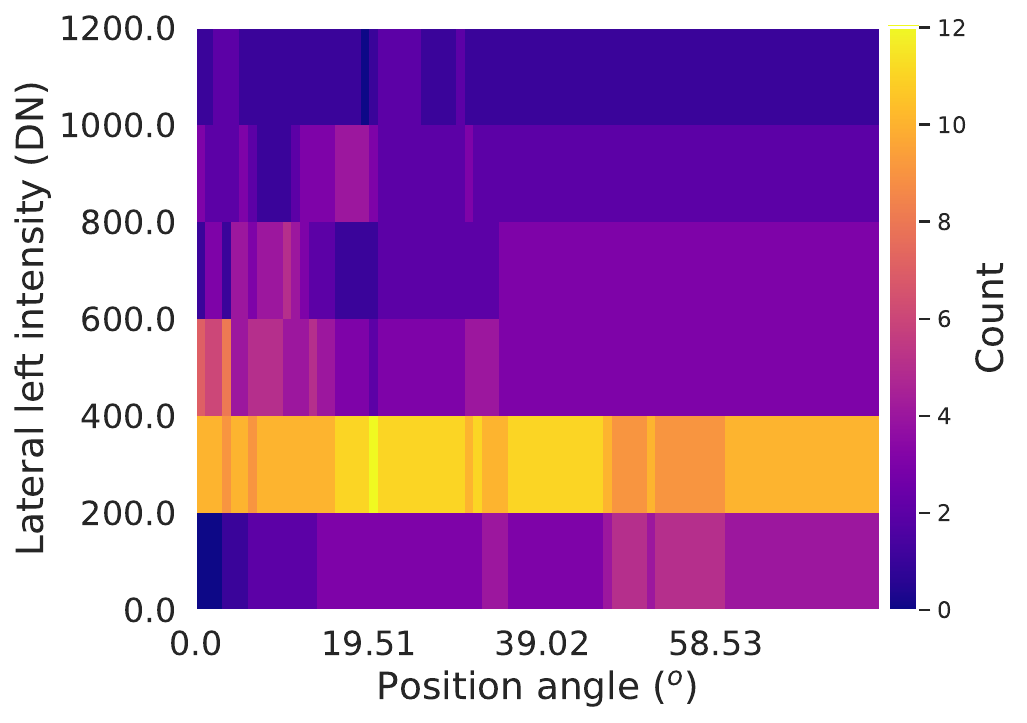}
      \includegraphics[width=0.32\textwidth,clip=]{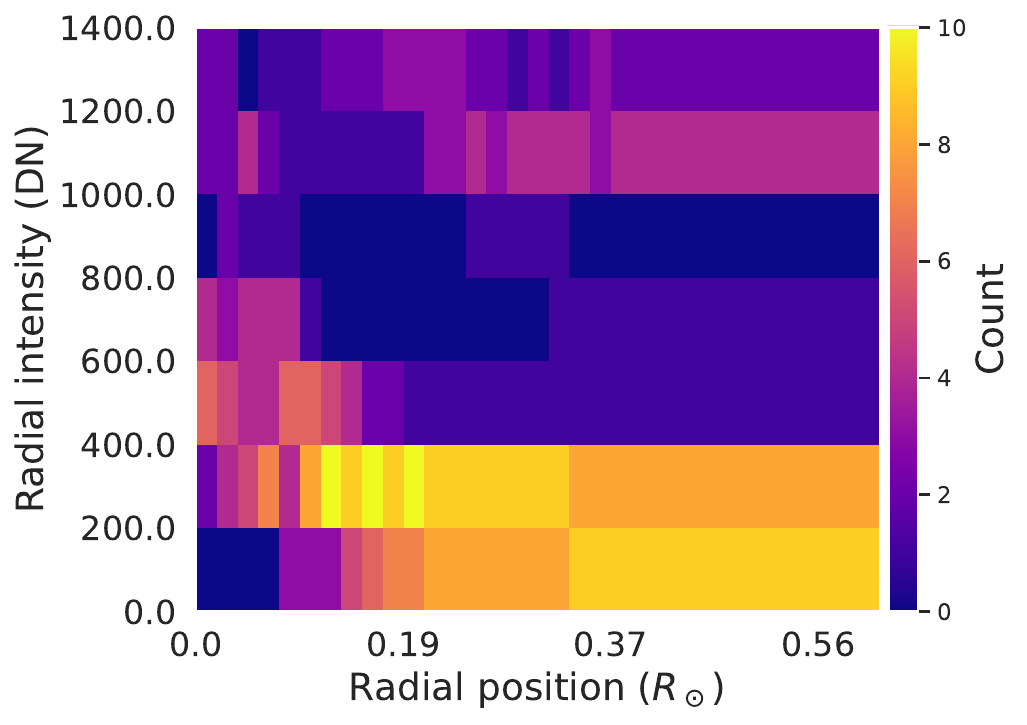}
      \includegraphics[width=0.32\textwidth,clip=]{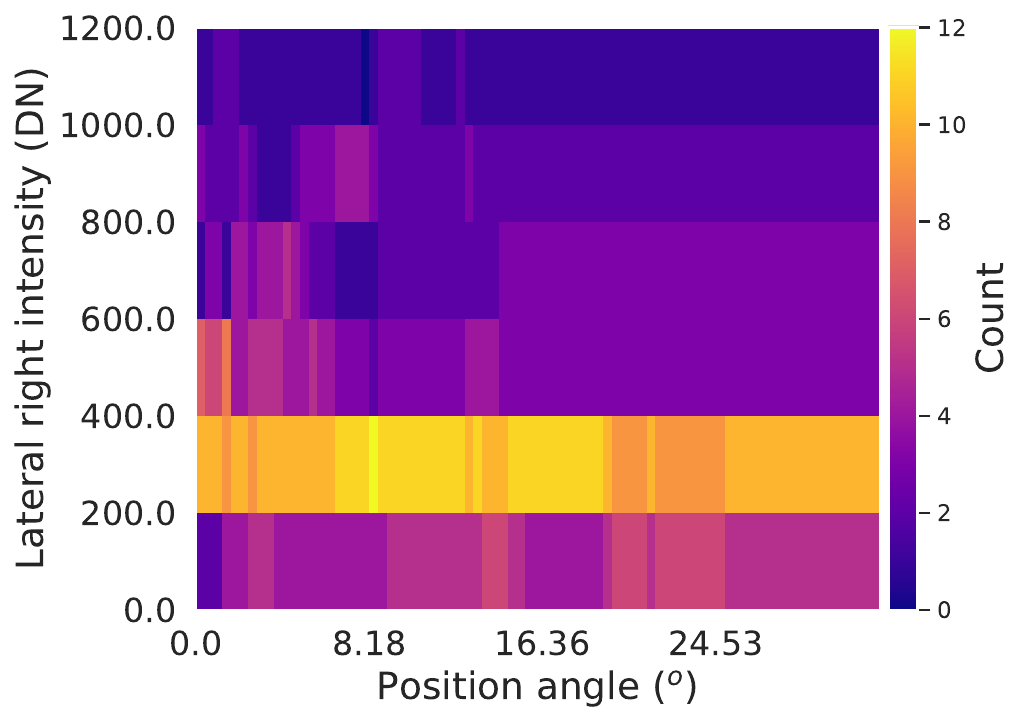}
      }
  \centerline{
      \includegraphics[width=0.32\textwidth,clip=]{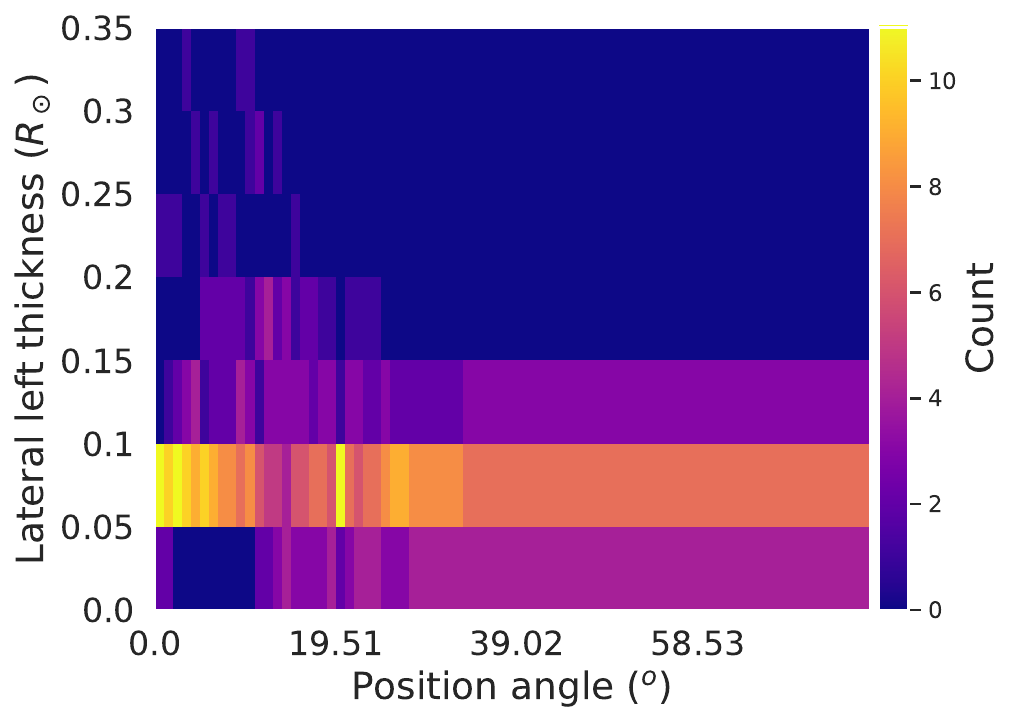}
      \includegraphics[width=0.32\textwidth,clip=]{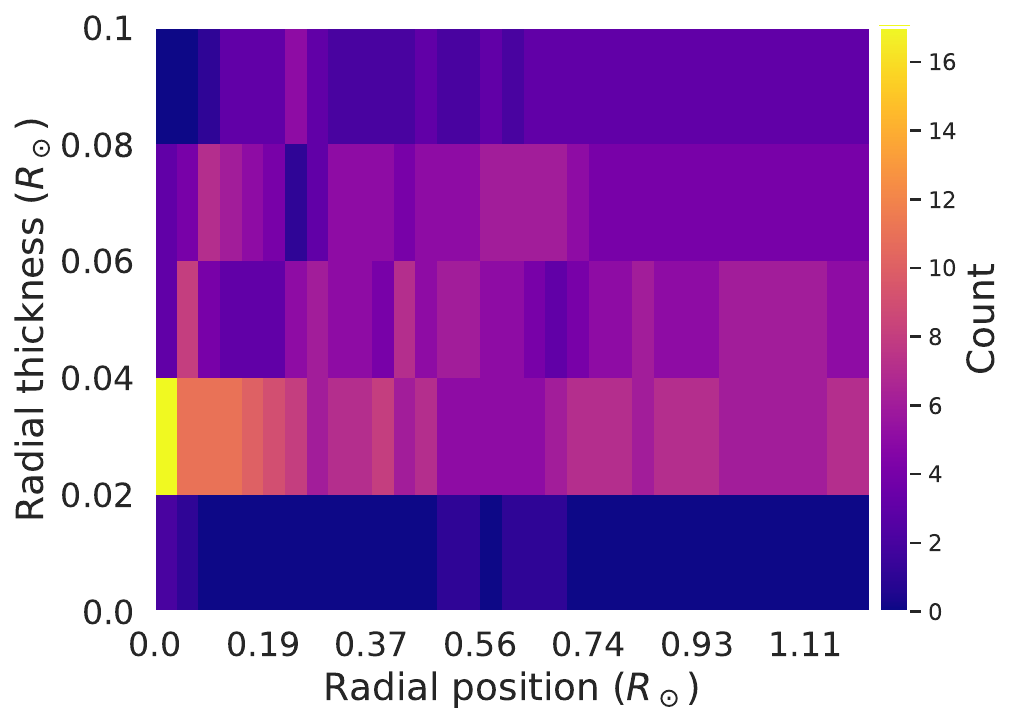}
      \includegraphics[width=0.32\textwidth,clip=]{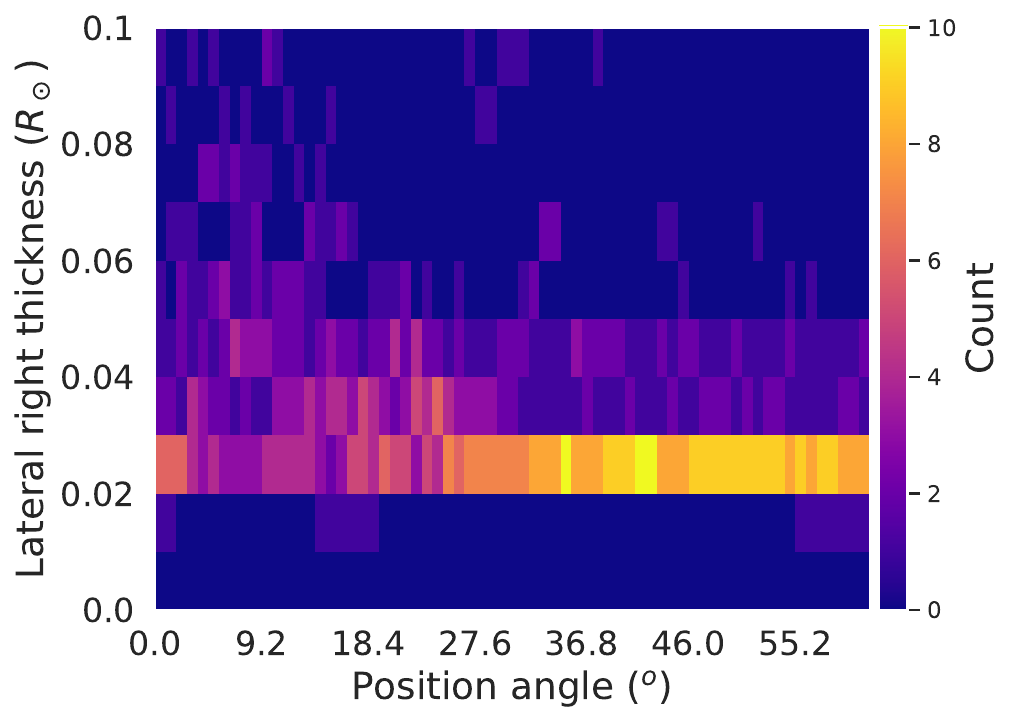}
      }
  \caption{Dynamic spectra of the EUV waves kinematics in the AIA FOV. The panels from the top to the bottom are the wave speeds, acceleration, mean intensity, and thickness. The left column is for the lateral left flank, the central column is for the radial direction, and the right column is for the lateral right flank.}
  \label{fig_kinematics_spect_hist}
\end{figure}

To illustrate the evolution of EUV waves' kinematics in the AIA FOV, we present Figure~\ref{fig_kinematics_spect_hist}, which provides a cumulative view of dynamic spectra for all events. The figure showcases the parameter distribution as a function of distance for shock speed, acceleration, wave intensity, and wave thickness for the radial direction (the middle column) and the lateral directions; the left and right flanks in left and right columns, respectively. The colors in the figure represent the total count in each bin at each radial position step, or each position angle step.

Consistent with our expectations, the speed and intensity panels exhibit a decline in values as a function of distance. As the waves propagate away from the Sun, the wave drivers lose momentum through interactions with the medium, leading to a decrease in speed. Additionally, plasma densities decrease with distance, resulting in a corresponding decrease in wave intensity.

\begin{table}[!htp] % updated!
\centering
\caption{Statistics of the EUV wave kinematics in the SDO/AIA FOV for the 26 events. LL and LR refer to the lateral left and right flanks, respectively. Rad refer to the radial front direction.}
\label{T_sh_param_all}
\resizebox{\textwidth}{!}{%
\begin{tabular}{lccccccccccccc}
\hline
       &              & \multicolumn{3}{c}{Speed (\kms)}  & \multicolumn{3}{c}{Accel. (km s$^{-2}$)} & \multicolumn{3}{c}{Intensity (DN)} & \multicolumn{3}{c}{Thickness (\rsun)} \\ \hline
       & Aspect ratio & LL      & Rad     & LR     & LL      & Rad     & LR     & LL       & Rad      & LR      & LL       & Rad      & LR      \\ \hline
Max    & 2.00         & 1574.81 & 2053.73 & 983.58 & 28.19   & 81.01   & 13.89  & 1348.87  & 2431.95  & 1498.45 & 9.600    & 0.185    & 6.100   \\ \hline
Min    & 0.84         & 2.11    & 40.30   & 2.30   & -35.24  & -81.01  & -9.89  & 0.53     & 0.17     & 150.30  & 0.027    & 0.018    & 0.022   \\ \hline
Mean   & 1.87         & 316.17  & 413.60  & 264.50 & -0.15   & 0.98    & 0.13   & 438.99   & 681.46   & 442.46  & 0.715    & 0.059    & 0.231   \\ \hline
Median & 2.00         & 284.77  & 349.32  & 216.32 & 0.03    & 0.37    & 0.11   & 337.96   & 425.23   & 389.06  & 0.102    & 0.055    & 0.076   \\ \hline
Stdv.  & 0.33         & 261.01  & 336.11  & 191.13 & 5.53    & 11.08   & 2.05   & 292.26   & 592.78   & 227.10  & 1.721    & 0.030    & 0.776   \\ \hline
\end{tabular}%
}
\end{table}

\begin{figure}[!htp] % updated!
  \centerline{\includegraphics[width=0.7\columnwidth]{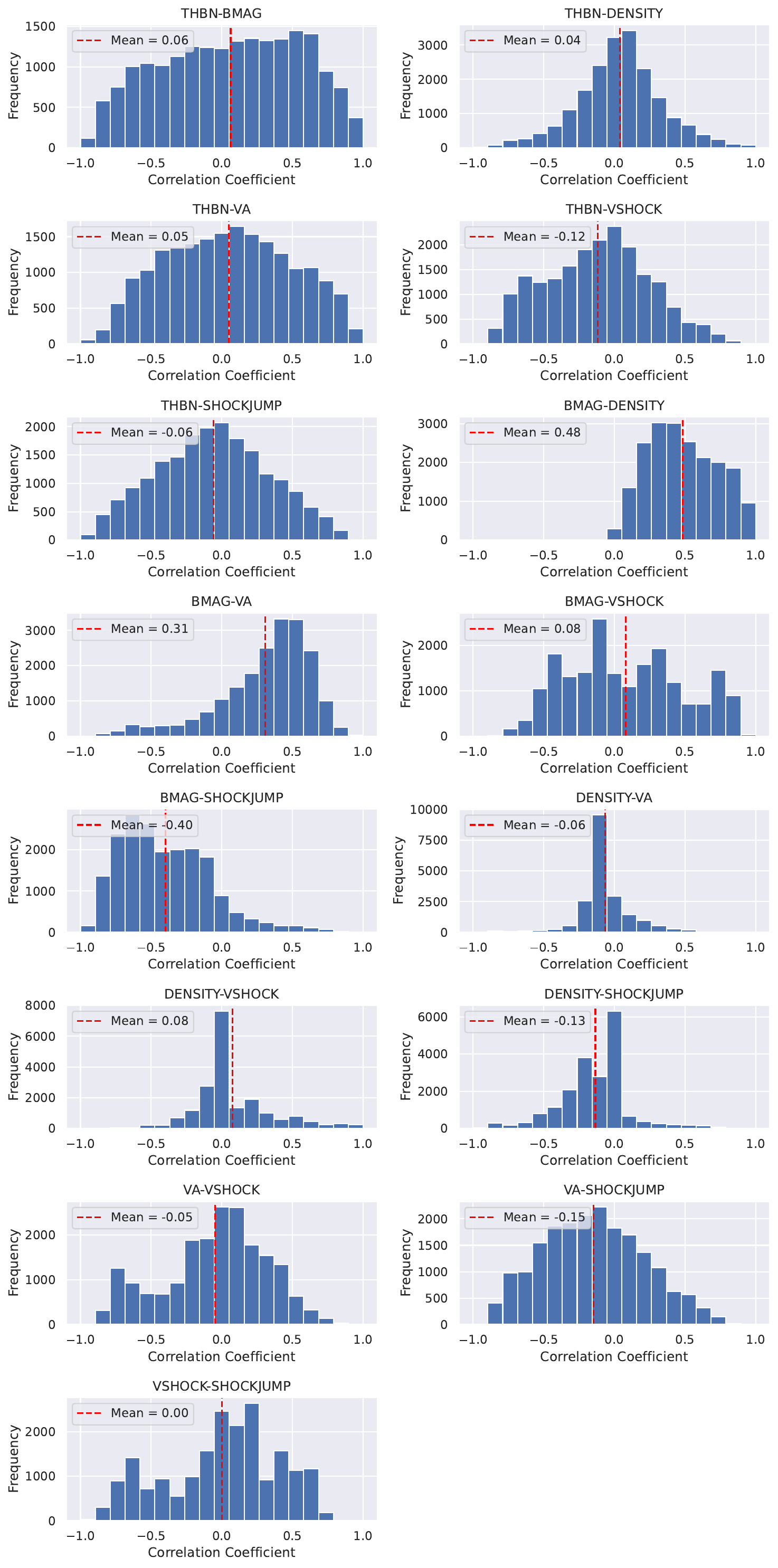}}
  \caption{Histograms of along-field-lines model plasma parameters in the solar corona for all the 26 events. The vertical dashed red lines are the mean values.}
  \label{fig_hist_plasma_param_corr}
\end{figure}

To investigate the bulk behavior of the modeled plasma above and at the shock surface, we sampled over 1000 field lines from the 26 events. The resulting histograms, shown in Figures~\ref{fig_hist_plasma_param_corr}, reveal correlations between various pairs of parameters. 
We investigated the correlations between five parameters -- the shock-field angle (\textbf{THBN}), the coronal magnetic field (\textbf{BMAG}), the plasma density (\textbf{DENSITY}), the Alfven speed (\textbf{VA}), the shock speed (\textbf{VSHOCK}), and the shock density jump (\textbf{SHOCKJUMP}).

The histograms show that there are weak to moderate correlations between some of the parameters.
For example, there is a moderate positive correlation between \textit{BMAG} and \textit{DENSITY}, and a moderate positive correlation between \textit{BMAG} and \textit{VA}. These correlations suggest that there may be some underlying physical processes that connect these two parameters.
The positive correlation between \textit{BMAG} and \textit{DENSITY} could be due to the fact that stronger magnetic fields can compress the plasma, leading to higher densities.
The negative correlation between \textit{BMAG} and \textit{VA} could be due to the fact that stronger magnetic fields tend to speed up the Alfven waves.

In addition to the anticipated correlations between the Alfven speed with magnetic field and density, we discovered a highly skewed correlation between magnetic field values and the modeled shock density jump, as well as between magnetic field magnitude and density.
The negative correlation between \textit{BMAG} and \textit{SHOCKJUMP} indicates that stronger coronal magnetic fields are associated with smaller density jumps across the shock surface. In other words, weak magnetic field correlates well with stronger shocks.
This could be due to several possible mechanisms. For instance, stronger magnetic fields exert higher pressure, potentially resisting the compression of plasma by the shock wave, leading to a smaller density increase across the shock front.
In addition, as mentioned earlier, stronger magnetic fields might lead to faster Alfven waves, allowing the plasma ahead of the shock to react and reduce the density jump.
These correlations will be further explored to establish a more definitive connection and to parameterize the shock density jump.

We also aim to investigate the event-averaged modeled plasma parameters and establish the observed connections between these parameters throughout entire events. To achieve this, we computed the mean, maximum, and summed values of the parameters for each event. The analysis revealed a set of parameter pairs that exhibit promising relationships suitable for parameterization. Among these pairs, we selected six representative cases for further examination.

\begin{figure}[!htp] % updated!
  \centerline{
    \includegraphics[width=0.45\textwidth]{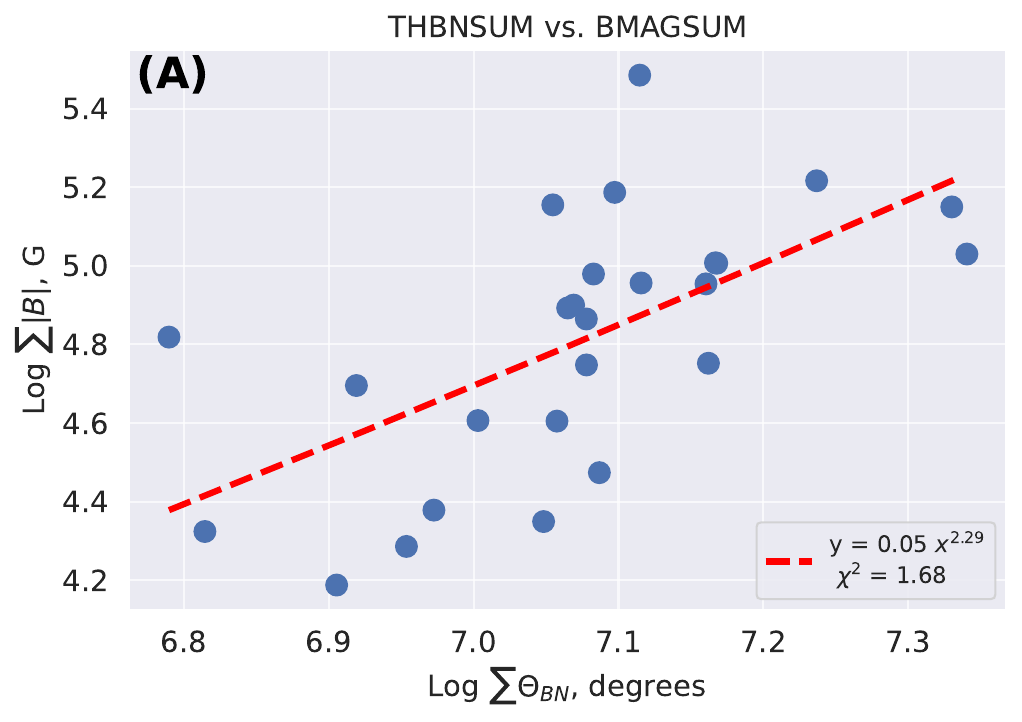}
    \includegraphics[width=0.45\textwidth]{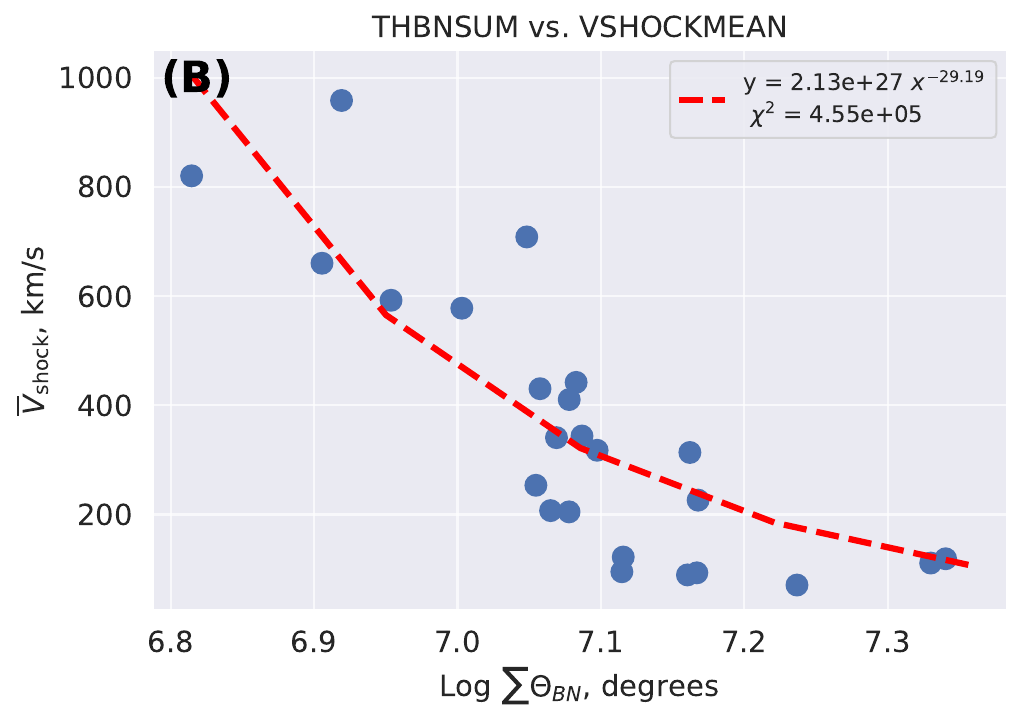}
    }
  \centerline{
    \includegraphics[width=0.45\textwidth]{figs/THBNSUM_VSHOCKSUM.pdf}
    \includegraphics[width=0.45\textwidth]{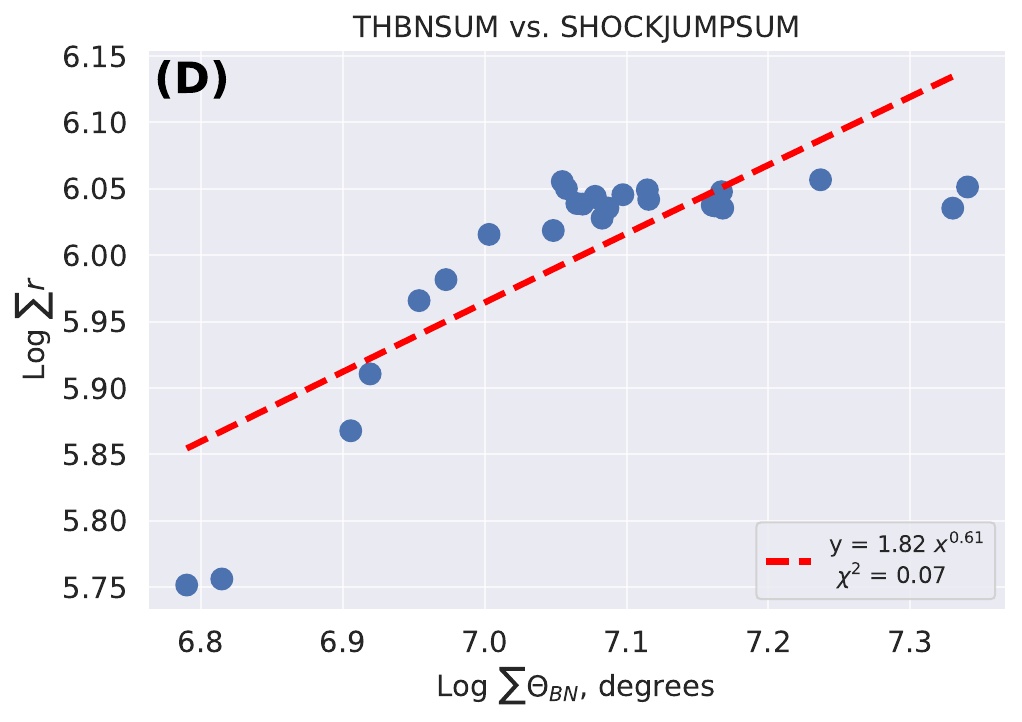}
    }
  \centerline{
    \includegraphics[width=0.45\textwidth]{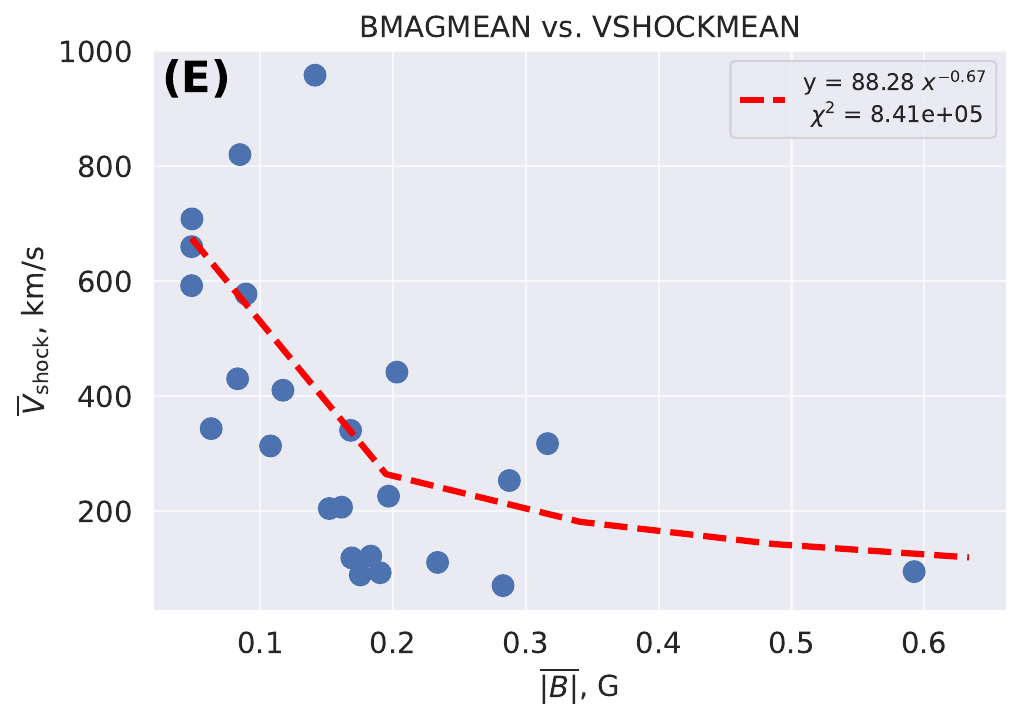}
    \includegraphics[width=0.45\textwidth]{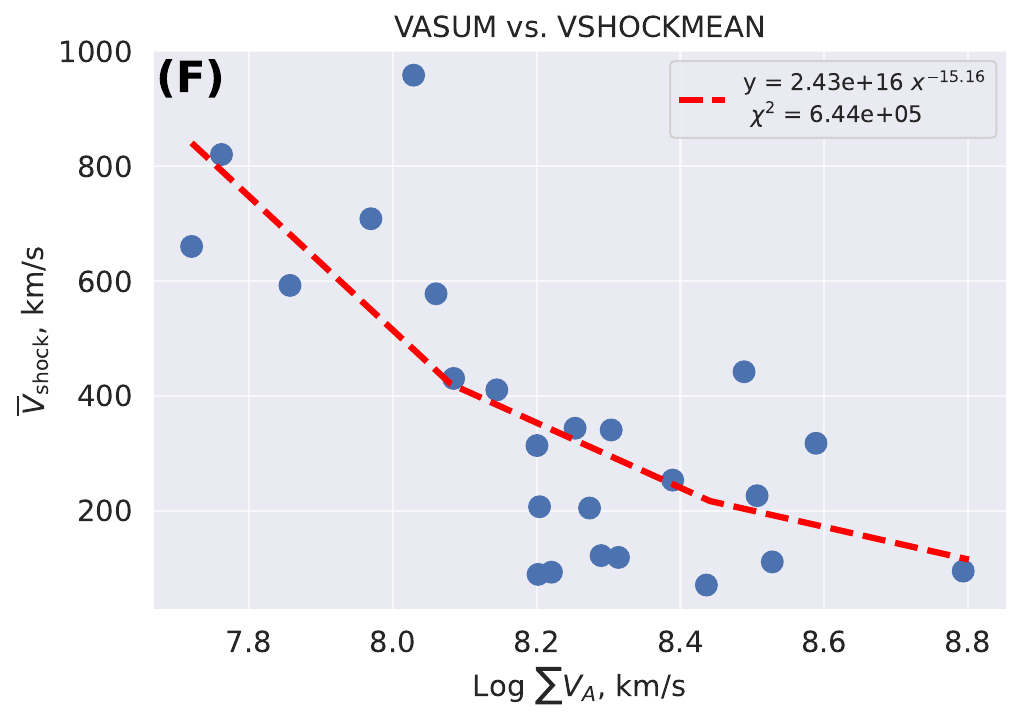}
    }
  \centerline{
    \includegraphics[width=0.45\textwidth]{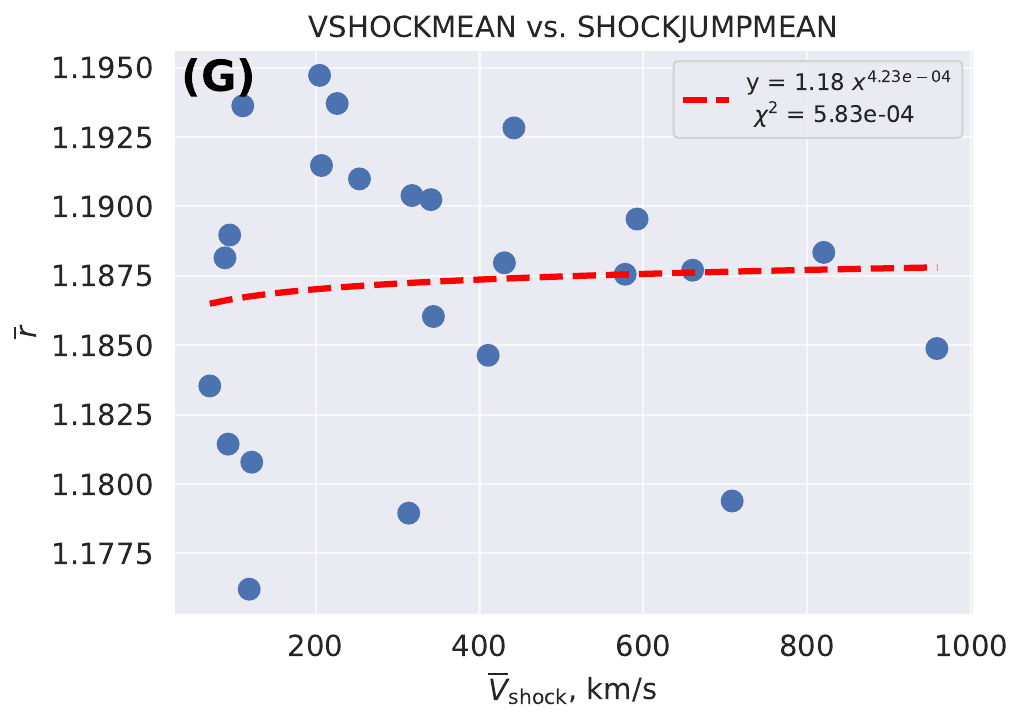}
    \includegraphics[width=0.45\textwidth]{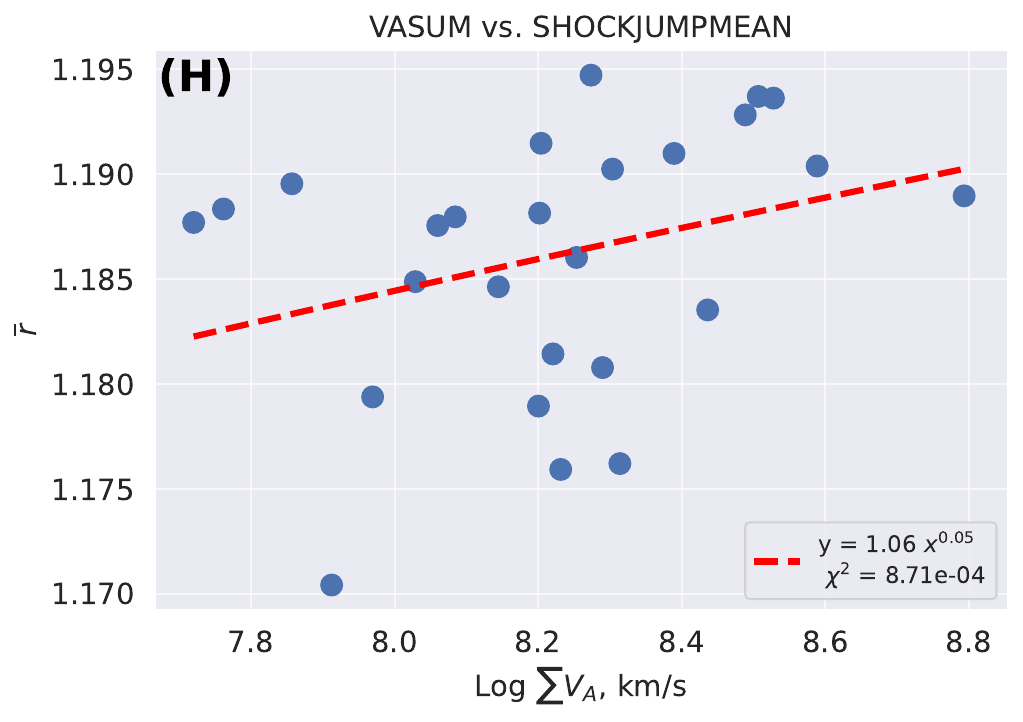}
    }
  \caption{Scatter plots of 8 coronal plasma-parameter pairs that exhibit parameterizable relationships. The VSHOCKMEAN is filtered to take only events with speeds $<$4000 \kms.}
  \label{fig_scatterplot_plasma_param}
\end{figure}

The scatter plots presented in Figure~\ref{fig_scatterplot_plasma_param} depict the chosen modeled parameter pairs. From left to right, top to the bottom, we plotted the logarithm of the summed values of the shock-field angle $\theta_{BN}$ versus the logarithm of the summed values of the absolute coronal magnetic field (A), the logarithm of the summed values of $\theta_{BN}$ versus the mean values of the shock speed (B), the logarithm of the summed values of $\theta_{BN}$ versus the logarithm of the summed values of the shock speed (C), the logarithm of the summed values of $\theta_{BN}$ versus the logarithm of the summed values of the shock density jump (D), the mean values of the absolute magnetic field versus the mean values of the shock speed (E), the logarithm of the summed values of the Alfven speed versus the mean values of the shock speed (F), the mean values of the shock speed versus the mean values of the density jump (G), and finally the logarithm of the summed values of the Alfven speed versus the mean values of the density jump (H).

To determine the best-fitting equations, we tested several models and found that power fits yield the most accurate results in this particular case. The graphs illustrating the fit parameters and the $\chi^2$ are provided for reference.

The analysis of eight figures examining shock dynamics in EUV wave events revealed significant findings. In Figure (A), a weak positive correlation between coronal magnetic field strength and shock-field angle was observed, suggesting a nuanced relationship influenced by factors beyond magnetic fields. Figure (B) showcased a weak negative correlation between shock-field angles and mean shock speeds, hinting at a connection between complex shock geometries and slower wave speeds. In Figure (C), there appears to be a weak negative correlation between the logarithm of the shock-field angle and the logarithm of the sum of shock speeds, which is indicated by the faintly downward-sloping trendline. This hints at a slight tendency for larger shock-field angles to be associated with slower shock speeds. Figure (D) displayed a moderate positive correlation between shock density jump and shock-field angles, suggesting potential interactions between shock wave geometry and density variations.

Meanwhile, Figures (E), (F), (G), and (H) delved into relationships between solar coronal magnetic field strength, shock speeds, Alfven speeds, and shock density jumps, revealing negative, lack of clear, and lack of consistent correlations, respectively. These findings emphasize the intricate nature of solar coronal shock dynamics, with multiple influencing factors contributing to observed correlations and scattering in the data. The comprehensive analyses underscore the need for more events and a nuanced understanding of various factors when interpreting the complexities of shock phenomena in the solar corona.

It is important to note that some outliers exist, but they do not undermine the overall correlation patterns.
In future work, we will focus on developing and testing parameterizations for these identified connections in order to establish a set of synoptic MHD parameters that correspond one-to-one with the parameters we measure, such as shock density jump and shock speed. By accomplishing this, we will enhance the representation of shock parameters even in the absence of actual compressive waves.

%______________________________________________________________ 
\section{Conclusions}
\label{s_conclusions}

We have conducted a comprehensive study focusing on the characterization of 26 historical CME-driven CBFs in the low solar corona. These events were accompanied by type III radio bursts, and SEP events near Earth and were observed by the AIA instrument onboard the SDO spacecraft in the EUV 193 $\AA$ band. To achieve this, we utilized the SPREAdFAST framework, which encompasses physics-based and data-driven models to estimate the coronal magnetic field, dynamics of coronal shock waves, energetic particle acceleration, and SEP propagation in the heliosphere.

Our analysis relied on sequences of base-difference images obtained from the AIA instrument. These images served as the primary input data for our investigation. We employed these data to generate annulas plots and J-maps to estimate the kinematic measurements in both the radial and lateral directions for the EUV waves.

In order to obtain a thorough understanding of the CBFs, we computed various time-dependent and distance-dependent kinematic parameters. These included shock speed, acceleration, intensity, and thickness of the front, peak, and back edges of the waves, as well as the major and minor axes and the rate of change of the waves' aspect ratios. To augment our analysis based on AIA observations, we incorporated LASCO measurements up to 17\rsun. This additional data is important in improving the characterization of the SEP spectra near the Sun.

The analysis of kinematic measurements played a pivotal role in our study as they enabled us to generate time-dependent 3D geometric models of wavefronts. In addition, these measurements provided valuable insights for deriving time-dependent plasma diagnostics by incorporating the outcomes of the MHD and DEM models.

To accurately represent the shocks, we employed shock kinematic measurements to fit a geometric spheroid surface model for each measured time step. This approach allowed us to capture the intricate characteristics of the shocks with precision.

In order to gain a deeper understanding of the phenomenon, we explored the parametrized relationships between the modeled plasma parameters. Through this analysis, we aimed to identify potential connections and interdependencies, shedding light on the complex dynamics at play.

Overall, our findings in this study and in \cite{kozarev_2022} as well as \cite{stepanyuk_2022} contribute to a nuanced understanding of shock kinematics and shock plasma parameters.

Moving forward, our future investigations will focus on examining SEP acceleration near the Sun, as well as investigating the transport of coronal and interplanetary particles using the insights gained from our models. Additionally, we aim to refine the methods of shock and coronal parameters characterization, which will contribute to enhancing the accuracy and reliability of the results.
%%________________________________________________________________

\section*{Acknowledgments}
This project is funded by an ESA Contract 4000126128 to the Institute of Astronomy and NAO, BAS, under the Plan for European Cooperating States (PECS) program in Bulgaria. This work is supported by the Bulgarian National Science Fund, VIHREN program, under contract KP-06-DV-8/18.12.2019 (MOSAIICS project). The authors acknowledge using data from the Atmospheric Imaging Assembly instrument onboard the Solar Dynamic Observatory, and data from the SOHO/LASCO CME catalog.

%\newpage
\bibliographystyle{apalike}
\bibliography{refs}

\end{document}